\documentclass[preprint,3p]{elsarticle} 

\usepackage{amssymb}
\usepackage{amsthm}
\usepackage{amsmath}
\usepackage{parskip}
\usepackage{graphicx}
\usepackage{algorithm}
\usepackage{color}
\usepackage{tikz}
\usepackage{multirow}
\usepackage{qtree}
\usepackage{url}

\usepackage{algorithm}
\usepackage[noend]{algorithmic}

\journal{Knowledge-based systems}
\begin{document}
\begin{frontmatter}



\newtheorem{definition}{Definition}
\newtheorem{theorem}{Theorem}[section]
\newtheorem{lemma}[theorem]{Lemma}
\newtheorem{proposition}[theorem]{Proposition}
\newtheorem{corollary}[theorem]{Corollary}

%

\hyphenation{}

\title{Web Data Extraction, Applications and Techniques: A Survey}


\author[1]{Emilio Ferrara\corref{cor1}}
\address[1]{Center for Complex Networks and Systems Research, Indiana University, Bloomington, IN 47408, USA}
\ead{ferrarae@indiana.edu}
\cortext[cor1]{Corresponding author}
\author[2]{Pasquale De Meo}
\address[2]{Univ. of Messina, Dept. of Ancient and Modern Civilization, Polo Annunziata, I-98166 Messina, Italy}
\ead{pdemeo@unime.it}
\author[3]{Giacomo Fiumara}
\address[3]{Univ. of Messina, Dept. of Mathematics and Informatics, viale F. Stagno D'Alcontres 31, I-98166 Messina, Italy}
\ead{gfiumara@unime.it}
\author[4]{Robert Baumgartner}
\address[4]{Lixto Software GmbH, Austria}
\ead{robert.baumgartner@lixto.com}


\address{}

\begin{abstract}

Web Data Extraction is an important problem that has been studied by means of different scientific tools and in a broad range of applications. Many approaches to extracting data from the Web have been designed to solve specific problems and operate in ad-hoc domains. Other approaches, instead, heavily reuse techniques and algorithms developed in the field of Information Extraction.

This survey aims at providing a structured and comprehensive overview of the literature in the field of Web Data Extraction. 
We provided a simple classification framework in which existing Web Data Extraction applications are grouped into two main classes, namely applications at the Enterprise level and at the Social Web level. At the Enterprise level, Web Data Extraction techniques emerge as a key tool to perform data analysis in Business and Competitive Intelligence systems as well as for business process re-engineering. At the Social Web level, Web Data Extraction techniques allow to gather a large amount of structured data continuously generated and disseminated by Web 2.0, Social Media and Online Social Network users and this offers unprecedented opportunities to analyze human behavior at a very large scale.
We discuss also the potential of cross-fertilization, i.e., on the possibility of re-using Web Data Extraction techniques originally designed to work in a given domain, in other domains.
\end{abstract}


\end{frontmatter}


\newpage

\tableofcontents

\newpage

\section{Introduction}
\label{sec:intro}

Web Data Extraction systems are a broad class of software applications targeting at extracting information from Web sources \cite{1,8}.
A Web Data Extraction system usually interacts with a Web source and extracts data stored in it: for instance, if the source is an HTML Web page, the extracted information could consist of elements in the page as well as the full-text of the page itself.
Eventually, extracted data might be post-processed, converted in the most convenient structured format and stored for further usage \cite{32,34}.

Web Data Extraction systems find extensive use in a wide range of applications including the analysis of text-based documents available to a company (like e-mails, support forums, technical and legal documentation, and so on), Business and Competitive Intelligence \cite{45}, crawling of Social Web platforms \cite{Catanese2011,Gjoka2010}, Bio-Informatics \cite{14} and so on.
The importance of Web Data Extraction systems depends on the fact that a large (and steadily growing) amount of information is continuously produced, shared and consumed online: Web Data Extraction systems allow to efficiently collect this information with limited human effort.
The availability and analysis of collected data is an indefeasible requirement to understand complex social, scientific and economic phenomena which generate the information itself.
For example, collecting digital traces produced by users of Social Web platforms like Facebook, YouTube or Flickr is the key step to understand, model and predict human behavior \cite{kleinberg2000smallworld,newman2003structure,backstrom2011four}.

In the commercial field, the Web provides a wealth of public domain information.
A company can probe the Web to acquire and analyze information about the activity of its competitors. This process is known as {\em Competitive Intelligence} \cite{46,47} and it is crucial to quickly identify the opportunities provided by the market, to anticipate the decisions of the competitors as well as to learn from their faults and successes.

\subsection{Challenges of Web Data Extraction techniques}
\label{sub:challenges}

The design and implementation of Web Data Extraction systems has been discussed from different perspectives and it leverages on scientific methods coming from  various disciplines including Machine Learning, Logic and Natural Language Processing.

In the design of a Web Data Extraction system, many factors must be taken into account; some of them are independent of the specific application domain in which we plan to perform Web Data Extraction.
Other factors, instead, heavily depend on the particular features of the application domain: as a consequence, some technological solutions which appear to be effective in some application contexts are not suitable in others.

In its most general formulation, the problem of extracting data from the Web is hard because it is constrained by several requirements.
The key challenges we can encounter in the design of a Web Data Extraction system can be summarized as follows:

\begin{itemize}

\item Web Data Extraction techniques often require the help of human experts.
A first challenge consists of providing a {\em high degree of automation} by reducing human efforts as much as possible.
Human feedback, however, may play an important role in raising the level of accuracy achieved by a Web Data Extraction system.

A related challenge is, therefore, to identify a reasonable trade-off between the need of building highly automated Web Data Extraction procedures and the requirement of achieving accurate performance.

\item Web Data Extraction techniques should be able to process large volumes of data in relatively short time.
This requirement is particularly stringent in the field of Business and Competitive Intelligence because a company needs to perform timely analysis of market conditions.

\item Applications in the field of Social Web or, more in general, those dealing with personal data must provide solid privacy guarantees.
Therefore, potential (even if unintentional) attempts to violate user privacy should be timely and adequately identified and counteracted.

\item Approaches relying on Machine Learning often require a significantly large training set of manually labeled Web pages.
In general, the task of labeling pages is time-expensive and error-prone and, therefore, in many cases we can not assume the existence of labeled pages.

\item Oftentimes, a Web Data Extraction tool has to routinely extract data from a Web Data source which can evolve over time.
Web sources are continuously evolving and structural changes happen with no forewarning, thus are unpredictable.
Eventually, in real-world scenarios it emerges the need of maintaining these systems, that might stop working correctly if lacking of flexibility to detect and face structural modifications of related Web sources.

\end{itemize}

\subsection{Related work}
\label{sub:related}

The theme of Web Data Extraction is covered by a number of reviews.
Laender et al. \cite{1} presented a survey that offers a rigorous \textit{taxonomy} to classify Web Data Extraction systems. The authors introduced a set of criteria and a qualitative analysis of various Web Data Extraction tools.

Kushmerick \cite{2} defined a profile of \textit{finite-state approaches} to the Web Data Extraction problem.
The author analyzed both wrapper induction approaches (i.e., approaches capable of automatically generating wrappers by exploiting suitable examples) and maintenance ones (i.e., methods to update a wrapper each time the structure of the Web source changes).
In that paper, Web Data Extraction techniques derived from Natural Language Processing and Hidden Markov Models were also discussed.
On the wrapper induction problem, Flesca et al. \cite{3} and Kaiser and Miksch \cite{4} surveyed approaches, techniques and tools.
The latter paper, in particular, provided a model describing the architecture of an Information Extraction system.
Chang et al. \cite{5} introduced a \textit{tri-dimensional categorization} of Web Data Extraction systems, based on task difficulties, techniques used and degree of automation.
In 2007, Fiumara \cite{6} applied these criteria to classify four state-of-the-art Web Data Extraction systems.
A relevant survey on Information Extraction is due to Sarawagi  \cite{7} and, in our opinion, anybody who intends to approach this discipline should read it.
Recently, some authors focused on {\em unstructured data management systems (UDMSs)} \cite{doan2009information}, i.e., software systems that analyze raw text data, extract from them some structure (e.g. person name and location), integrate the structure (e.g., objects like {\em New York} and {\em NYC} are merged into a single object) and use the integrated structure to build a database.
UDMSs are a relevant example of Web Data Extraction systems and the work from Doan et al. \cite{doan2009information} provides an overview of {\tt Cimple}, an UDMS developed at the University of Wisconsin.
To the best of our knowledge, the survey from Baumgartner et al. \cite{8} is the most recently updated review on the discipline as of this work.

\subsection{Our contribution} \label{sub:contribution}

The goal of this survey is to provide a structured and comprehensive overview of the research in Web Data Extraction as well as to provide an overview of most recent results in the literature.

We adopt a different point of view with respect to that used in other survey on this discipline: most of them present a list of tools, reporting a feature-based classification or an experimental comparison of these tools.
Many of these papers are solid starting points in the study of this area.
Unlike the existing surveys, our ambition is to provide a classification of existing Web Data Extraction techniques in terms of the application domains in which they have been employed.
We want to shed light on the various research directions in this field as well as to understand to what extent techniques initially applied in a particular application domain have been later re-used in others.
To the best of our knowledge, this is the first survey that deeply analyzes Web Data Extraction systems from a perspective of their application fields.

However, we also provide a detailed discussion of techniques to perform Web Data Extraction.
We identify two main categories, i.e., approaches based on Tree Matching algorithms and approaches based on Machine Learning algorithms.
For each category, we first describe the basic employed techniques and then we illustrate their variants.
We also show how each category addresses the problems of wrapper generation and maintenance.
After that, we focus on applications that are strictly interconnected with Web Data Extraction tasks.
We cover in particular enterprise, social and scientific applications by discussing which fields have already been approached (e.g., advertising engineering, enterprise solutions, Business and Competitive intelligence, etc.) and which are potentially going to be in the future (e.g., Bio-informatics, Web Harvesting, etc.).

We also discuss about the potential of {\em cross-fertilization}, i.e., whether strategies employed in a given domain can be re-used in others or, otherwise, if some applications can be adopted only in particular domains.

\subsection{Organization of the survey}

This survey is organized into two main parts.
The first one is devoted to provide general definitions which are helpful to understand the material proposed in the survey.
To this purpose, Section \ref{sec:techniques} illustrates the techniques exploited for collecting data from Web sources, and the algorithms that underlay most of Web Data Extraction systems.
The main features of existing Web Data Extraction systems are largely discussed in Section \ref{sec:webdataextraction}.

The second part of this work is about the applications of Web Data Extraction systems to real-world scenarios.
In Section \ref{sec:applications} we identify two main domains in which Web Data Extraction techniques have been employed: applications at the enterprise level and at the Social Web level. The formers are described in Section \ref{sec:enterprise}, whereas the latters are covered in Section \ref{sec:social-applications}.
This part concludes discussing the opportunities of cross-fertilization among different application scenarios (see Section \ref{sec:cross}).

In Section \ref{sec:conclusions} we draw our conclusions and discuss potential applications of Web Data Extraction techniques that might arise in the future.

\section{Techniques}
\label{sec:techniques}

The first part of this survey is devoted to the discussion of the techniques adopted in the field of the Web Data Extraction.
In this part we extensively review approaches to extracting data from HTML pages. HTML is the predominant language for implementing Web pages and it is largely supported by W3C consortium. HTML pages can be regarded as a form of semi-structured data (even if less structured than other sources like XML documents) in which information follows a nested structure; HTML features can be profitably used in the design of suitable wrappers.
However, we acknowledge that a large amount of semi-structured information is present in non-HTML formats (think of e-mail messages, software code and related documentations, system logs and so on) but the research approaches targeting at extracting information from this type of sources are out of the scope of this work.

The first attempts to extract data from the Web are dated back in early nineties.
In the early days, this discipline borrowed approaches and techniques from Information Extraction (IE) literature.
In particular, two classes of strategies emerged \cite{4}: \textit{learning techniques} and \textit{knowledge
engineering techniques} -- also called \textit{learning-based} and \textit{rule-based} approaches, respectively \cite{7}.
These classes share a common rationale: the former was thought to develop systems that require human expertise to define rules (for example, \textit{regular expressions}) to successfully accomplish the data extraction.
These approaches require specific domain expertise: users that design and implement the rules and train the system must have programming experience and a good knowledge of the domain in which the data extraction system will operate; they will also have the ability to envisage potential usage scenarios and tasks assigned to the system.
On the other hand, also some approaches of the latter class involve strong familiarity with both the requirements and the functions of the platform, so the human engagement is essential.

Various strategies have been devised to reduce the commitment of human domain experts.
Some of them have been developed in the context of Artificial Intelligence literature, like the adoption of specific algorithms that use the structure of Web pages to identify and extract data.
Others methods are borrowed from Machine Learning, like supervised or semi-supervised learning techniques to design systems capable of being trained by examples and then able to autonomously extract data from similar (or even different) domains.

In the following we will discuss separately these two research lines.
Section \ref{sub:wrappers} presents the strategies based on the definition of algorithms capable of identifying information exploiting the semi-structured nature of Web pages.
In Section \ref{sub:webwrappers} we introduce the concept of Web wrappers and explain how the techniques discussed in Section \ref{sub:wrappers} are incorporated into Web wrappers; the then  introduce some techniques for their generation and maintenance.
Section \ref{sub:hybrid-systems} poses the attention on a new class of platforms that sit between approaches based on machine learning to induce wrappers and platforms to generate wrappers.

\subsection{Tree-based techniques} \label{sub:wrappers}
One of the most exploited features in Web Data Extraction is the semi-structured nature of Web pages.
These can be naturally represented as \emph{labeled ordered rooted trees}, where labels represent the tags proper of the HTML mark-up language syntax, and the tree hierarchy represents the different levels of nesting of elements constituting the Web page.
The representation of a Web page by using a labeled ordered rooted tree is usually referred as \emph{DOM (Document Object Model)}, whose detailed explanation is out of the scope of this survey but has been largely regulated by the World Wide Web Consortium -- For further details consider the following link: {\tt www.w3.org/DOM}.
The general idea behind the Document Object Model is that HTML Web pages are represented by means of plain text that contains HTML tags, particular keywords defined in the mark-up language, which can be interpreted by the browser to represent the elements specific of a Web page (e.g., hyper-links, buttons, images and so forth), so as free-text.
HTML tags may be nested one into another, forming a hierarchical structure.
This hierarchy is captured in the DOM by the document tree whose nodes represent HTML tags.
The document tree (henceforth also referred as DOM tree) has been successfully exploited for Web Data Extraction purposes in a number of techniques discussed in the following.

\subsubsection{Addressing elements in the document tree: XPath} \label{sub:xpath}
One of the main advantage of the adoption of the Document Object Model for the HTML language is the possibility of exploiting some tools typical of XML languages (and HTML is to all effects a dialect of the XML).
In particular, the XML Path Language (or, briefly, XPath) provides with a powerful syntax to address specific elements of an XML document (and, to the same extent, of HTML Web pages) in a simple manner.
XPath has been defined by the World Wide Web Consortium, so as DOM -- For the specifications see: {\tt www.w3.org/TR/xpath}.

\begin{figure}
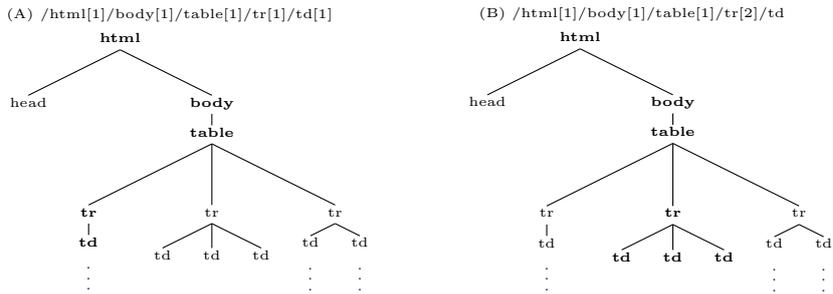
%
	\centering
	\tiny
	
	\begin{minipage}{6cm}
		\begin{center}
		\tiny (A) /html[1]/body[1]/table[1]/tr[1]/td[1]
		\smallskip
		
		\Tree [.{\bf html} [.head ] [.{\bf body} [.{\bf table} [.{\bf tr} [.{\bf td\\$\vdots$} ] ] [.tr [.td ] [.td ] [.td ] ] [.tr [.{td\\$\vdots$} ] [.{td\\$\vdots$} ] ] ]  ] ]
				
		\end{center}		
	\end{minipage}
	\begin{minipage}{6cm}
		\begin{center}
		\tiny (B) /html[1]/body[1]/table[1]/tr[2]/td
		\smallskip
		
		\Tree [.{\bf html} [.head ] [.{\bf body} [.{\bf table} [.tr [.{td\\$\vdots$} ] ] [.{\bf tr} [.{\bf td} ] [.{\bf td} ] [.{\bf td} ] ] [.tr [.{td\\$\vdots$} ] [.{td\\$\vdots$} ] ] ]  ] ]
	\end{center}
	
	\end{minipage}
	
	\caption{Example of XPath(s) on the document tree, selecting one (A) or multiple (B) items.}%
	\label{fig:xpath}%
\end{figure}

Although describing the syntax of XPath is not the core argument of this section, we provide Figure \ref{fig:xpath} as an example to explain how XPath can be used to address elements of a Web page.
There exist two possible ways to use XPath: \emph{(i)} to identify a single element in the document tree, or \emph{(ii)} to address multiple occurrences of the same element.
In the former case, illustrated in Figure \ref{fig:xpath}(A), the defined XPath identifies just a single element on the Web page (namely, a table cell); in the latter, showed in Figure \ref{fig:xpath}(B), the XPath identifies multiple instances of the same type of element (still a table cell) sharing the same hierarchical location.

To the purpose of Web Data Extraction, the possibility of exploiting such a powerful tool has been of utmost importance: the adoption of XPath as the tool to address elements in Web page has been largely exploited in the literature.
The major weakness of XPath is related to its lack of flexibility: each XPath expression is strictly related to the structure of the Web page on top of which it has been defined. However, this limitation has been partially mitigated, by introducing \emph{relative path expressions}, in latest releases (see XPath 2.0: {\tt www.w3.org/TR/xpath20}).
In general, even minor changes to the structure of a Web page might corrupt the correct functioning of an XPath expression defined on a previous version of the page.

To better clarify this concept, let us consider Web pages generated by a script (e.g., think of the information about a book in an e-commerce Web site).
Now assume that the script undergoes some change: we can expect that the tree structure of the HTML page generated by that script will change accordingly.
To keep the Web Data Extraction process functional, one should update the expression every time any change occurs to the underlying page generation model; such an operation would require a high human commitment and, therefore, its cost could be prohibitively large.
To this purpose, some authors \cite{dalvi2009robust,dalvi2011automatic} introduced the concept of {\em wrapper robustness}: they proposed a strategy to find, among all the XPath expressions capable of extracting the same information from a Web page, the one that is less influenced by potential changes in the structure of the page and such an expression identifies the more robust wrapper. In general, to make the whole Web Data Extraction process robust, we need suitable tools allowing us to measure the similarity degree of two documents; such a task can be accomplished by detecting structural variations in the DOM trees associated with the documents. To this extent, some techniques, called {\em tree-matching strategies} are a good candidate to detect similarities between two tree and they will be discussed in detail in the next sections.

\subsubsection{Tree edit distance matching algorithms} \label{sub:tree-edit-distance}
The first technique we describe is called \emph{tree edit distance matching}.
The problem of computing the tree edit distance between trees is a variation of the classic string edit distance problem.
Given two labeled ordered rooted trees $A$ and $B$, the problem is to finding a matching to transform $A$ in $B$ (or the vice-versa) with the minimum number of operations.
The set of possible operations consists of node deletion, insertion or replacement.
At each operation might be applied a cost, and in that case, the task turns in a cost-minimization problem (i.e., finding the sequence of operations of minimum cost to transform $A$ in $B$).

The reasoning above is formally encoded in the definition of \emph{mapping}, presented by \cite{liu2011structured}.
A mapping $M$ between two trees $A$ and $B$ is defined as a \emph{set of ordered pairs} ($i$, $j$), one from each tree, satisfying the following conditions $\forall$ ($i_1$, $j_1$), ($i_2$, $j_2$) $\in M$.

\begin{enumerate}
 	\item $i_1 = i_2$ if and only if $j_1 = j_2$;
 	\item $A[i_1]$ is on the left of $A[i_2]$ if and only if $B[i_1]$ is on the left of $B[i_2]$;
 	\item $A[i_1]$ is an ancestor of $A[i_2]$ if and only if $B[i_1]$ is an ancestor of $B[i_2]$.
\end{enumerate}

With the notation $A[i_x]$ we indicate the x-\emph{th} node of the tree $A$ in a \emph{pre-order visit} of the tree.
A number of consequences emerge from this definition of mapping:

\begin{itemize}
 	\item Each node must appear no more than once in a mapping;
 	\item The order among siblings is preserved;
 	\item The hierarchical relations among nodes is unchanged.
\end{itemize}

A number of techniques to approach this problem have been proposed \cite{wang1998algorithm,chen2001new}. These methods support all three types of operations on nodes (i.e., node deletion, insertion and replacement) but are plagued by high computational costs.
It has also been proved that the formulation for non ordered trees is NP-complete \cite{zhang1992editing}.

\paragraph*{The simple tree matching algorithm}
A computationally efficient solution for the problem of the tree edit distance matching is provided by the algorithm called \emph{simple tree matching} \cite{selkow1977tree}, and its variants.
This optimized strategy comes at a cost: node replacement is not allowed during the matching procedure -- the shortcomings of this aspects will be further discussed below.
The pseudo-code of the simple tree matching algorithm is provided in Algorithm \ref{alg:simple-tree-matching}, which adopts the following notation: \emph{d(n)} represents the degree of a node \emph{n} (i.e., the number of first-level children); $T(i)$ is the i-\emph{th} subtree of the tree rooted at node $T$.

\begin{algorithm}[H]
\caption{SimpleTreeMatching($T^{'}$, $T^{''}$)}
\label{alg1}
\begin{algorithmic}[1]
    \IF{$T^{'}$ has the same label of $T^{''}$}
        \STATE $m \leftarrow$ $d(T^{'})$
        \STATE $n \leftarrow$ $d(T^{''})$
        \FOR{$i = 0$ to $m$}
            \STATE $M[i][0] \leftarrow 0$;
        \ENDFOR
        \FOR{$j = 0$ to $n$}
            \STATE $M[0][j] \leftarrow 0$;
        \ENDFOR
        \FORALL{$i$ such that $1\leq i\leq m$}
            \FORALL{$j$ such that $1\leq j \leq n$}
                \STATE $M[i][j] \leftarrow$ Max($M[i][j-1]$, $M[i-1][j]$, $M[i-1][j-1] + W[i][j]$) where $W[i][j]$ = SimpleTreeMatching($T^{'}(i-1)$, $T^{''}(j-1)$)
            \ENDFOR
        \ENDFOR
        \STATE return M[m][n]+1
    \ELSE
        \STATE return 0
    \ENDIF
\end{algorithmic}
\label{alg:simple-tree-matching}
\end{algorithm}

The computational cost of simple tree matching is $\mathsf{O(nodes(A) \cdot nodes(B)})$, where $nodes(T)$ is the function that returns the number of nodes in a tree (or a sub-tree) $T$; the low cost ensures excellent performance when applied to HTML trees, which might be rich of nodes.
Two main limitations hold with this algorithm:
\begin{itemize}
 	\item It can not match permutation of nodes;
 	\item No level crossing is allowed (it is impossible to match nodes at different hierarchical levels).
\end{itemize}

Despite these intrinsic limits, this technique appears to fit very well to the purpose of matching HTML trees in the context of Web Data Extraction systems.
In fact, it has been adopted in several scenarios \cite{kim2007web,19,ZhaiL06,zhang2010blog,69,ferrara2011design,ferrara2011intelligent}.
One of the first Web Data Extraction approaches based on a tree edit distance algorithm is due to Reis {\em et al.} \cite{reis2004automatic}.
Such an approach focused on a very specific application domain (i.e, news extraction from the Web) but it was general enough to be re-applied in other domains.
The algorithm of Reis {\em et al.} relies on a different definition of mapping called Restricted Top-Down Mapping (RTDM).
In this model, insertion, removal and replacement operations are allowed only to the leaves of the trees.
The restricted top-down edit distance between two trees $A$ and $B$ is defined as the cost of the restricted top-down mapping between the two trees.
To find the restricted top-down mapping between two trees $A$ and $B$, the Yang's algorithm is applied \cite{yang1991identifying}.
The worst case time complexity of the approach of \cite{reis2004automatic} is still $\mathsf{O(nodes(A) \cdot nodes(B)})$ but, as shown by the authors, it works much faster because only restricted top-down mappings are managed.
A normalized variant of the simple tree matching is called \emph{normalized simple tree matching}.
The normalization is computed as follows
$$
NSTM(A, B) = \frac{SimpleTreeMatching(A, B)}{(nodes(A) + nodes(B)) / 2}.
$$
The tree Matching algorithm and its variants are widely used in practice because they are easy to implement.

\paragraph*{The weighted tree matching algorithm} \label{sub:weighted-tree}
Another variant of the simple tree matching is discussed in the following, and is called \emph{weighted tree matching}.
It adjusts the similarity values provided by the original simple tree matching by introducing a re-normalization factor.
The pseudo-codify of the weighted tree matching, recently presented in \cite{69}, is reported in Algorithm \ref{alg2}.

\begin{algorithm}[H]
\caption{WeightedTreeMatching($T^{'}$, $T^{''}$)}
\label{alg2}
\begin{algorithmic}[1]
    \STATE \COMMENT{Change line 11 in Algorithm \ref{alg1} with the following code}
    \IF{$m > 0$ AND $n > 0$}
        \STATE return M[m][n] * 1 / Max($t(T^{'})$, $t(T^{''})$)
    \ELSE
        \STATE return M[m][n] + 1 / Max($t(T^{'})$, $t(T^{''})$)
    \ENDIF
\end{algorithmic}
\end{algorithm}

In Algorithm \ref{alg2}, the notation  $t(n)$ represents the number of total siblings of a node \emph{n} including itself.
Note that Algorithm \ref{alg2} reports the differential version with respect to the simple tree matching  described in Algorithm \ref{alg1}.
The advantage of the weighted tree matching is that it better reflects a measure of similarity between two trees.
In fact, in the simple tree matching algorithm the assigned matching value is always equal to one.
Instead, the weighted tree matching algorithm assumes that less importance (i.e., a lower weight) is assigned to changes in the structure of the tree, when they occur in deeper sub-levels.
This kind of changes can be, for example, missing or added leaves, truncated or added branches, etc.
Also, a lower weight is accounted when changes occur in sub-levels with many nodes.
The \emph{weighted tree matching} algorithm returns a value in the interval [0,1] and the closer to 1 the final value, the more similar the two input trees.

\begin{figure}
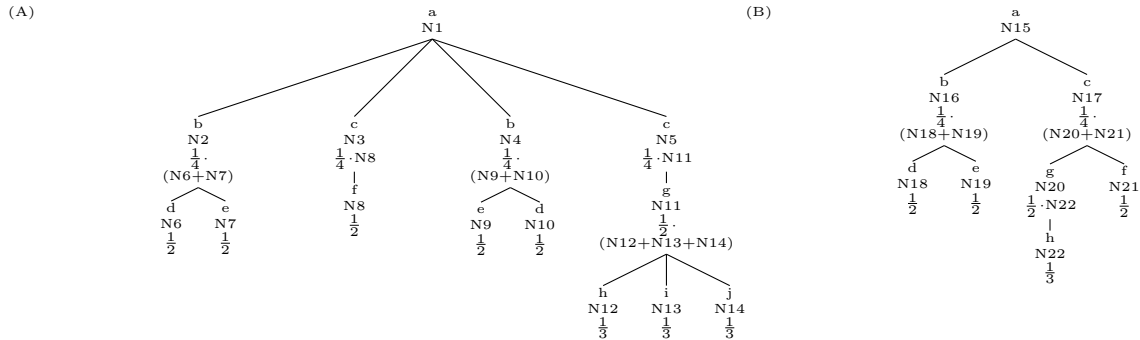

\tiny (A)
\Tree [.{a\\\tiny N1} [.{b\\\tiny N2\\$\frac{1}{4}\cdot$\\\tiny (N6+N7)} [.{d\\\tiny N6\\ $\frac{1}{2}$} ] [.{e\\\tiny N7\\ $\frac{1}{2}$} ] ] [.{c\\\tiny N3\\$\frac{1}{4}\cdot$\tiny N8} [.{f\\\tiny N8\\ $\frac{1}{2}$} ]  ] [.{b\\\tiny N4\\$\frac{1}{4}\cdot$\\\tiny (N9+N10)} [.{e\\\tiny N9\\ $\frac{1}{2}$} ] [.{d\\\tiny N10\\ $\frac{1}{2}$} ] ] [.{c\\\tiny N5\\$\frac{1}{4}\cdot$\tiny N11} [.{g\\\tiny N11\\$\frac{1}{2}\cdot$\\\tiny (N12+N13+N14)} [.{h\\\tiny N12\\ $\frac{1}{3}$} ] [.{i\\\tiny N13\\ $\frac{1}{3}$} ] [.{j\\\tiny N14\\ $\frac{1}{3}$} ] ] ] ]
\tiny (B)
\Tree [.{a\\\tiny N15} [.{b\\\tiny N16\\$\frac{1}{4}\cdot$\\\tiny (N18+N19)} [.{d\\\tiny N18\\ $\frac{1}{2}$} ] [.{e\\\tiny N19\\ $\frac{1}{2}$} ]  ] [.{c\\\tiny N17\\$\frac{1}{4}\cdot$\\\tiny (N20+N21)} [.{g\\\tiny N20\\$\frac{1}{2}\cdot$\tiny N22} [.{h\\\tiny N22\\ $\frac{1}{3}$} ] ] [.{f\\\tiny N21\\ $\frac{1}{2}$} ] ] ]
\caption{Example of application of the weighted tree matching algorithm for the comparison of two labeled rooted trees, \emph{A} and \emph{B}.}
\label{ex2}

\end{figure}

Let us analyze the behavior of the algorithm with an example often used in the literature \cite{yang1991identifying,19,69} to explain the simple tree matching (see Figure \ref{ex2}).
In that figure, $A$ and $B$ are two very simple generic rooted labeled trees (i.e., the same structure of HTML document trees).
They show several similarities except for some missing nodes/branches.
By applying the \emph{weighted tree matching} algorithm, a value of $\frac{1}{3}$ is established for nodes (h), (i) and (j) belonging to $A$, although two of them are missing in $B$.
Going up to parents, the summation of contributions of matching leaves is multiplied by the relative value of each node (e.g., in the first sub-level, the contribution of each node is $\frac{1}{4}$ because of the four first-sublevel nodes in A).
Once completed these operations for all nodes of the sub-level, values are added and the final measure of similarity for the two trees is obtained.

In this example, the weighted tree matching between $A$ and $B$ returns a measure of similarity of $\frac{3}{8}$ ($0.375$) whereas the simple tree matching would return 7.
The main difference on results provided by these two algorithms is the following: the \emph{weighted tree matching} intrinsically produces a proper measure of similarity between the two compared trees while the \emph{simple tree matching} returns the mapping value.
A remarkable feature of the \emph{weighted tree matching} algorithm is that, the more the structure of considered trees is complex and similar, the more the measure of similarity will be accurate.


\subsection{Web wrappers}
\label{sub:webwrappers}

In the previous section we discussed some algorithms that might be adopted to identify information exploiting the semi-structured format of HTML documents.
In the following we will discuss those procedures that might adopt the techniques presented above to carry out the data extraction.

In the literature, any procedure that aims at extracting structure data from unstructured (or semi-structured) data sources is usually referred as \emph{wrapper}.
In the context of Web Data Extraction we provide the following definition:

\begin{definition}[Web wrapper]
A procedure, that might implement one or many different classes of algorithms, which \textit{seeks} and \textit{finds} data required by a human user, extracting them from unstructured (or semi-structured) Web sources, and \textit{transforming} them into structured data, \textit{merging} and \textit{unifying} this information for \textit{further processing}, in a semi-automatic or fully automatic way.
\end{definition}

Web wrappers are characterized by a life-cycle constituted by several steps:

\begin{enumerate}
 	\item Wrapper generation: the wrapper is defined according to some technique(s);
 	\item Wrapper execution: the wrapper runs and extracts information continuously;
 	\item Wrapper maintenance: the structure of data sources may change and the wrapper should be adapted accordingly to keep working properly.
\end{enumerate}

In the remainder of this section we will discuss these three different phases.

In particular, the first two steps of a wrapper life-cycle, its generation and execution, are discussed in Section \ref{wrapper:generation}; these steps might be implemented \textit{manually}, for example by defining and executing regular expressions over the HTML documents; alternatively, which is the aim of Web Data Extraction systems, wrappers might be defined and executed by using an inductive approach -- a process commonly known as \textit{wrapper induction} \cite{38}.
Web wrapper induction is challenging because it requires high level automation strategies.
There exist also hybrid approaches that make possible for users to generate and run wrappers semi-automatically by means of visual interfaces.

The last step of a wrapper life-cycle is the maintenance: Web pages change their structure continuously and without forewarning.
This might corrupt the correct functioning of a Web wrapper, whose definition is usually tightly bound to the structure of the Web pages adopted for its generation.
Defining automatic strategies for wrapper maintenance is of outstanding importance to guarantee correctness of extracted data and robustness of Web Data Extraction platforms.
Some methodologies have been recently presented in the literature, as discussed in Section \ref{wrapper:maintenance}.


\subsubsection{Wrapper generation and execution} \label{wrapper:generation}

The first step in wrappers life-cycle is their generation.
Early Web Data Extraction platforms provided only support for manual generation of wrappers, which required human expertise and skills in programming languages to write scripts able to identify and extract selected pieces of information within a Web page.

In late nineties they made their appearance more advanced Web Data Extraction systems.
The core feature provided was the possibility for their users to define and execute Web wrappers by means of interactive graphical users interfaces (GUIs).
In most cases, it was not required any deep understanding of a wrapper programming language, as wrappers were generated automatically (or semi-automatically) by the system exploiting directives given by users by means of the platform interface.

In the following we discuss in detail three types of rationales underlying these kind of platforms, namely \emph{regular expressions}, \emph{wrapper programming languages} and \emph{tree-based approaches}.
The details regarding the features and of different Web Data Extraction platforms, instead, will be described in great detail in Section \ref{sec:webdataextraction}.

\paragraph*{Regular-expression-based approach}
One of the most common approaches is based on regular expressions, which are a powerful formal language used to identify strings or patterns in unstructured text on the basis of some matching criteria.
Rules could be complex so, writing them manually, could require much time and a great expertise: wrappers based on regular expressions dynamically generate rules to extract desired data from Web pages.
Usually, writing regular expressions on HTML pages relies on the following criteria: word boundaries, HTML tags, tables structure, etc.
The advantage of platforms relying on regular expressions is that the user can usually select (for example by means of a graphical interface) one (or multiple) element(s) in a Web page, and the system is able to automatically infer the appropriate regular expression to identify that element in the page.
Then, a wrapper might be created so that to extract similar elements, from other Web pages with the same structure of the one adopted to infer the regular expressions.

A notable tool implementing regular-expression-based extraction is W4F \cite{11}.
W4F adopts an annotation approach: instead of challenging users to deal with the HTML documents syntax, W4F eases the design of the wrapper by means of a \emph{wizard procedure}.
This wizard allows users to select and annotate elements directly on the Web page.
W4F produces the regular expression extraction rules of the annotated items and provides them to users.
A further step, which is the optimization of the regular expressions generated by W4F, is delegated to expert users -- in fact, the tool is not always able to provide the best extraction rule.
By fully exploiting the power of regular expressions, W4F extraction rules include \textit{match} and also  \textit{split} expressions, which separates words, annotating different elements on the same string.
The drawback of the adoption of regular expressions is their lack of flexibility.
For example, whenever even a minor change occurs in the structure or content of a Web page, each regular expression is very likely to stop working, and must be rewritten.
This process implies a big commitment by human users, in particular in the maintenance of systems based on regular expressions.
For this reasons more flexible and powerful languages have been developed to empower the capabilities of Web Data Extraction platforms.

\paragraph*{Logic-based approach} \label{sub:logic-based}
One example of powerful languages developed for data extraction purposes comes from the Web specific \emph{wrapper programming languages}.
Tools based on wrapper programming languages consider Web pages not as simply text strings but as semi-structured tree documents, whereas the DOM of the Web page represents its structure where nodes are elements characterized by both their properties and their content.
The advantage of such an approach is that wrapper programming languages might be defined to fully exploit both the semi-structured nature of the Web pages and their contents -- the former aspect lacks in regular-expression-based systems.

The first powerful wrapping language has been formalized by Gottlob and Koch \cite{18}.
The information extraction functions implemented by this wrapping language rely on monadic datalogs over trees \cite{28}.
The authors demonstrated that monadic datalogs over tree are equivalent to monadic second order logic (MSO), and hence very expressive. However, unlike MSO, a wrapper in monadic datalogs can be modeled nicely in a visual and interactive step-by-step manner.
This makes this wrappring language suitable for being incorporated into visual tools, satisfying the condition that all its constructs can be implemented through \textit{corresponding visual primitives}.

A bit of flavor on the functioning of the wrapping language is provided in the following.
Starting from the unranked labeled tree representing the DOM of the Web page, the algorithm re-labels nodes, truncates the irrelevant ones, and finally returns a subset of original tree nodes, representing the selected data extracted.
The first implementation of this wrapping language in a real-world scenarios is due to Baumgartner et al. \cite{10,12}.
They developed the \emph{Elog} wrapping language that implements almost all monadic datalog information extraction functions, with some minor restrictions.
The Elog language is used as the core extraction method of the Lixto Visual Wrapper system.
This platform provides a GUI to select, through visual specification, patterns in Web pages, in hierarchical order, highlighting elements of the document and specifying relationships among them.
Information identified in this way could be too general, thus the system allows users to add some restricting conditions, for example before/after, not-before/not-after, internal and range conditions.
Finally, selected data are translated into XML documents, by using pattern names as XML element names, obtaining structured data from semi-structured Web pages.

\paragraph*{Tree-based approach [partial tree alignment]}
The last technique discussed in this part relates to wrapper generation and is called \emph{partial tree alignment}.
It has been recently formalized by Zhai and Liu \cite{19,ZhaiL06} and the authors also developed a Web Data Extraction system based on it.
This technique relies on the idea that information in Web documents usually are collected in contiguous regions of the page, called \textit{data record regions}.
The strategy of partial tree alignment consists in identifying and extracting these regions.
In particular, the authors take inspiration from tree matching algorithms, by using the already discussed \textit{tree edit distance} matching (see Section \ref{sub:tree-edit-distance}).
The algorithm works in two steps:

\begin{enumerate}
 	\item Segmentation;
 	\item Partial tree alignment.
\end{enumerate}

In the first phase, the Web page is split in segments, without extracting any data.
This pre-processing phase is instrumental to the latter step.
In fact, the system not only performs an analysis of the Web page document based on the DOM tree, but also relies on visual cues (like in the spatial reasoning technique, see Section \ref{sub:spatial-reasoning}), trying to identify gaps between data records.
This step is useful also because helps the process of extracting structural information from the HTML document, in that situations when the HTML syntax is abused, for example by using tabular structure instead of CSS to arrange the graphical aspect of the page.

In the second step, the partial tree alignment algorithm is applied to data records earlier identified.
Each data record is extracted from its DOM sub-tree position, constituting the root of a new single tree.
This, because each data record could be contained in more than one non-contiguous sub-tree in the original DOM tree.
The partial tree alignment approach implies the alignment of data fields with certainty, excluding those that can not be aligned, to ensure a high degree of precision.
During this process no data items are involved, because partial tree alignment works only on tree tags matching, represented as the minimum cost, in terms of operations (i.e., node removal, node insertion, node replacement), to transform one node into another one.
The drawback of this characteristic of the algorithm is that its recall performance (i.e., the ability of recovering all expected information) might decay in case of complex HTML document structures.
In addition, also in the case of the partial tree alignment, the functioning of this strategy is strictly related with the structure of the Web page at the time of the definition of the alignment.
This implies that the method is very sensitive even to small changes, that might compromise the functioning of the algorithm and the correct extraction of information.
Even in this approach, the problem of the maintenance arises with outstanding importance.

\paragraph*{Machine Learning approaches} Machine Learning techniques fit well to the purpose of extracting domain-specific information from Web sources, since they rely on training sessions during which a system acquires a domain expertise.
Machine Learning approaches require a training step in which domain experts provide some manually labeled Web pages, acquired from different websites but also in the same website.
Particular attention should be paid to providing examples of Web pages belonging to the same domain but exhibiting different structures.
This, because, even in the same domain scenario, templates usually adopted to generate dynamic contents Web pages, differ, and the system should be capable of learning how to extract information in these contexts.
Statistical Machine Learning systems were also developed, relying on conditional models \cite{36} or adaptive search \cite{49} as an alternative solution to human knowledge and interaction.
In the following we shortly describe some Web Data Extraction approaches relying on Machine Learning algorithms \cite{21,califf2003bottom,mooney1999relational,soderland1999learning,freitag2000machine,22,23}.

One of the early approaches is {\em WIEN} \cite{21}. WIEN was based on different \textit{inductive learning techniques} and it was capable of automatically labeling training pages, representing \emph{de facto} a hybrid system whose training process  implied low human engagement.
The flip side of the high automation of WIEN was the big number of limitations related to its inferential system: for example, the data extraction process was not capable of dealing with missing values -- a case that occurs on a frequent base and posed serious limitations on the adaptability of WIEN to real-world scenarios.

{\em Rapier} (Robust Automated Production of Information Extraction Rules) \cite{califf2003bottom,mooney1999relational} is a system designed to learn
rules for extracting information from documents and its main advantage is, perhaps, the capability of learning these rules directly from documents  without prior parsing or any post-processing.
Extraction rules are relatively simple and they make use of limited syntactic and semantic information.

On one hand, Rapier rules are flexible because they are not restricted to contain a fixed number of words but, on the other hand, it is hard to recognize what rules are actually useful to perform data extraction.
To this purpose, a learning algorithm has been developed to find effective rules and this algorithm is based on Inductive Logic Programming.

{\em WHISK} was introduced in \cite{soderland1999learning}. It relies on a supervised learning algorithm that generates rules for extracting information from text documents. WHISK is able to handle a wide range of text documents ranging from highly structured documents (like HTML pages) to free text. The extraction rules considered in WHISK can be regarded as a special type of regular expressions that have two components: the former specifies the context in which a phrase has to be considered relevant, the latter specifies the exact delimiters of the phrase to be extracted (i.e., the bounds of the text that has to be extracted).
Depending of the structure of a document, WHISK generates rule that rely on exactly one of the two components cited above. In particular,  in case of free text it uses context-based rules whereas in case of structured text it uses delimiters.
In addition, for all those documents whose structure lies between structured document and free text WHISK is able to use a combination of context-based and delimiter-based rules.
WHISK uses a supervised learning algorithm to induce novel rules from a set of hand-tagged instances.
In order to keep human effort limited, WHISK interleaves the learning of new rules and the annotation of new instances.
The learning/annotation process is, therefore, iterative and rather than presenting arbitrary instances, WHISK presents instances that are near to the examples that can be managed by the rules WHISK has learned so far.

SRV was proposed by Freytag \cite{freitag2000machine}. SRV takes as input a set of tagged documents and extracts some features describing the tokens that can be extracted from a document.
Features are classified into {\em simple} if they map a token onto a categorical value and {\em relational} if they map a token onto another token.
SRV is also able to manage features encoding the structural aspect of a document (e.g., if a token is a verb).
Extraction rules can be expressed on the basis of available features.
To construct new rules, SRV uses a Naive Bayes classifier in conjunction with a relational learner.

SoftMealy \cite{22} was the first wrapper induction system specifically designed to work in the Web Data Extraction context.
Relying on \textit{non-deterministic finite state automata} (also known as finite-state transducers (FST)), SoftMealy uses a bottom-up inductive learning approach to learn extraction rules.
During the training session the system acquires training pages represented as an automaton on all the possible permutations of Web pages: states represent extracted data, while state transitions represent extraction rules.
SoftMealy's main strength was its novel method of internal representation of the HTML documents.
In detail, during a pre-processing step, each considered Web page was encoded into tokens (defined according to a set of inferential rules).
Then, tokens were exploited to define separators, considered as invisible borderlines between two consecutive tokens.
Finally, the FST was fed by \emph{sequence of separators}, instead of raw HTML strings (as in WIEN), so that to match tokens with contextual rules (defined to characterize a set of individual separators) to determine the state transitions.
The advantages of SoftMealy with respect to WIEN are worth noting: in fact, the system was able to deal with a number of exception, such as missing values/attributes, multiple attribute values, variant attribute permutations and also with typos.

The last learning-based system discussed in this part is called STALKER \cite{23}.
It was a supervised learning system for wrapper induction sharing some similarities with SoftMealy.
The main difference between these two systems is the specification of relevant data: in STALKER, a set of tokens is manually positioned on the Web page, so that to identify information that the user intend to extract.
This aspect ensures the capability of STALKER of handling with empty values, hierarchical structures and non ordered items.
This system models a Web page content by means of hierarchical relations, represented by using a tree data structure called \emph{embedded catalog tree} (EC tree).
The root of the EC tree is populated the sequence of all tokens (whereas, STALKER considers as token any piece of text or HTML tag in the document).
Each child node is a sub-sequence of tokens inherited by its parent node.
This implies that each parent node is a super-sequence of tokens of its children.
The super-sequence is used, at each level of the hierarchy, to keep track of the content in the sub-levels of the EC tree.
The extraction of elements of interest for the user is achieved by inferring a set of extraction rules on the EC tree itself -- a typical example of extraction rule inferred by STALKER is the construct \texttt{SkipTo(T)}, a directive that indicates, during the extraction phase, to skip all tokens until the first occurrence of the token \texttt{T} is found.
The inference of extraction rules exploits the concept of landmarks, sequences of consecutive tokens adopted to locate the beginning and the end of a given item to extract.
STALKER is also able to define \emph{wildcards}, classes of generic tokens that are inclusive of more specific tokens.

\subsubsection{The problem of wrapper maintenance}
\label{wrapper:maintenance}

Wrapper generation, regardless the adopted technique, is one aspect of the problem of data extraction from Web sources.
On the other hand, \emph{wrapper maintenance} is equally important, so that Web Data Extraction platform may reach high levels of robustness and reliability, hand in hand with the level of automation and low level of human engagement.
In fact, differently from static documents, Web pages dynamically change and evolve, and their structure may change, sometimes with the consequence that previously defined wrappers are no longer able to successfully extract data.

In the light of these assumptions, one could argument that wrapper maintenance is a critical step of the Web Data Extraction process.
Even though, this aspect has not acquired lot of attention in the literature (much less than the problem of wrapper generation), unless latest years.
In the early stage, in fact, wrapper maintenance was performed manually: users that usually design Web wrappers, were updating or rewriting these wrappers every time the structure of a given Web page was modified.
The manual maintenance approach was fitting well for small problems, but becomes unsuitable if the pool of Web pages largely increases.
Since in the enterprise scenarios regular data extraction tasks might involve thousand (or even more) Web pages, dynamically generated and frequently updated, the manual wrapper maintenance is not anymore a feasible solution for real-world applications.

For these reasons, the problem of automatizing the wrapper maintenance has been faced in recent literature.
For example, the first effort in the direction of automatic wrapper maintenance has been presented by Kushmerick \cite{2}, who defined for first the concept of \textit{wrapper verification}.
The task of wrapper verification arises as a required step during wrapper execution, in which a Web Data Extraction system assess if defined Web wrappers work properly or, alternatively, their functioning is corrupted due to modifications to the structure of underlying pages.
Subsequently, the author discussed some techniques of semi-automatic wrapper maintenance, to handle simple problems.

The first method that tries to automatize the process of wrapper maintenance has been developed by Meng et al. \cite{20}, and it is called \emph{schema-guided wrapper maintenance}.
It relies on the definition of XML schemes during the phase of the wrapper generation, to be exploited for the maintenance during the execution step.
More recently, Ferrara and Baumgartner \cite{69,ferrara2011design,ferrara2011intelligent} developed a system of automatic wrapper adaptation (a kind of maintenance that occur to modify Web wrappers according to the new structure of the Web pages) relying on the analysis of structural similarities between different versions of the same Web page using a tree-edit algorithm.
In the following, we discuss these two strategies of wrapper maintenance.

\paragraph*{Schema-guided wrapper maintenance}
The first attempt to deal with the problem of wrapper maintenance providing a high level of automation has been presented by Meng et al. \cite{20}.
The authors developed SG-WRAM (Schema-Guided WRApper Maintenance), a strategy for Web Data Extraction that is built on top of the assumption, based on empirical observations, that changes in Web pages, even substantial, often  preserve:

\begin{itemize}

\item Syntactic features: syntactic characteristics of data items like data patterns, string lengths, etc., are mostly preserved.

\item Hyperlinks: HTML documents are often enriched by hyperlinks that are seldom removed in subsequent modifications of the Web pages.

\item Annotations: descriptive information representing the semantic meaning of a piece of information in its context is usually maintained.
\end{itemize}

In the light of these assumptions, the authors developed a Web Data Extraction system that, during the phase of wrapper generation, creates schemes which will be exploited during the phase of wrapper maintenance.
In detail, during the generation of wrappers, the user provides HTML documents and XML schemes, specifying a mapping between them.
Later, the system will generate extraction rules and then it will execute the wrappers to extract data, building an XML document according to the specified XML schema.
During the wrapper execution phase, an additional component is introduced in the pipeline of the data extraction process: the \emph{wrapper maintainer}.
The wrapper maintainer checks for any potential extraction issue and provides an automatic repairing protocol for wrappers which fail their extraction task because of modifications in the structure of related Web pages.
The repairing protocol might be successful, and in that case the data extraction continues, or it might fail -- in that case warnings and notifications rise.
The XML schemes are defined in the format of a DTD (Document Type Definition) and the HTML documents are represented as DOM trees, according to what explained in Section \ref{sub:wrappers}.
The SG-WRAM system builds corresponding mappings between them and generates extraction rules in the format of XQuery expressions -- For XQuery specifications see: {\tt www.w3.org/TR/xquery}.

\paragraph*{Automatic wrapper adaptation}

Another strategy for the automatic maintenance of Web wrappers has been recently presented in \cite{69,ferrara2011design,ferrara2011intelligent}.
In detail, it is a method of automatic wrapper adaptation that relies on the idea of comparing helpful structural information stored in the Web wrapper defined on the original version of the Web page, searching for similarities in the new version of the page, after any structural modification occurs.

The strategy works for different data extraction techniques implemented by the wrapping system.
For example, it has been tested by using both XPath (see Section \ref{sub:xpath}) and the Elog wrapping language (see Section \ref{sub:logic-based}).
In this strategy, elements are identified and represented as sub-trees of the DOM tree of the Web page, and can be exploited to find similarities between two different versions of the same document.
We discuss an example adopting XPath to address a single element in a Web page, as reported in the example in Figure \ref{fig:xpath}(A).

The rationale behind the automatic wrapper adaptation is to search for some elements, in the modified version of the Web page, that share structural similarities with the original one.
The evaluation of the similarity is done on the basis of comparable features (e.g., subtrees, attributes, etc.).
These elements are called \emph{candidates}: among them, the one showing the higher degree of similarity with the  element in the original page, is matched in the new version of the page.
The algorithm adopted to compute the matching among the DOM trees of the two HTML documents is the weighted tree matching, already presented in Section \ref{sub:weighted-tree}.
Further heuristics are adopted to assess the similarity of nodes, for example exploiting additional features exhibited by DOM elements.
In some cases, for example, their textual contents might be compared, according to string distance measures such as Jaro-Winkler \cite{winkler1999state} or bigrams \cite{collins1996new}, to take into account also the content similarity of two given nodes.

It is possible to extend the same approach in the case in which the XPath identifies multiple similar elements on the original page (e.g., an XPath selecting results of a search in a retail online shop, represented as table rows, divs or list items), as reported in Figure \ref{fig:xpath}(B).
In detail, it is possible to identify multiple elements sharing a similar structure in the new page, within a custom level of accuracy (e.g., establishing a threshold value of similarity).

The authors implemented this approach in a commercial tool -- Lixto, describing how the pipeline of the wrapper generation has been modified to allow the wrappers to automatically detect and adapt their functioning to structural modifications occurring to Web pages \cite{ferrara2011intelligent}.
In this context, the strategy proposed was to acquire structural information about those elements the original wrapper extracts, storing them directly inside the wrapper itself.
This is done, for example, generating \emph{signatures} representing the DOM sub-tree of extracted elements from the original Web page, stored as a tree diagram, or as a simple XML documents.
During the execution of Web wrappers, if any extraction issue occurs due to a structural modification of the page, the wrapper adaptation algorithm automatically starts and tries to adapt the wrapper to the new structure.

\subsection{Hybrid systems: learning-based wrapper generation}
\label{sub:hybrid-systems}

Wrapper generation systems and wrapper induction techniques discussed above differ essentially in two aspects:

\begin{enumerate}
	\item The degree of automation of the Web Data Extraction systems;
	\item The amount and the type of human engagement required during operation.
\end{enumerate}

The first point is related to the ability of the system to work in an autonomous way, ensuring sufficient standards of robustness and reliability, according to the requirements of users.
Regarding the second point, most of the wrapper induction systems need labeled examples provided during the training sessions, thus requiring human expert engagement for the manual labeling phase.
Wrapper generation systems, on the other hand, engage users into their maintenance, unless automatic techniques are employed, as those discussed in Section \ref{wrapper:maintenance}.

Interestingly, a new class of platforms has been discussed in recent literature, that adopts a hybrid approach that sits between learning-based wrapper induction systems and wrapper generation platforms.
The first example of this class of systems is given by RoadRunner \cite{9,33}, a template-based system that automatically generates templates to extract data by matching features from different pages in the same domain.
Another interesting approach is that of exploiting visual cues and spatial reasoning to identify elements in the Web pages with a Computer Vision oriented paradigm.
This part concludes with the discussion of these two systems.

\paragraph*{Template-based matching}
The first example of hybrid system is provided by RoadRunner \cite{9,33}.
This system might be considered as an interesting example of automatic wrapper generator.
The main strength of RoadRunner is that it is oriented to data-intensive websites based on templates or regular structures.
The system tackles the problem of data extraction exploiting both features used by wrapper generators, and by wrapper induction systems.
In particular, RoadRunner can work using information provided by users, in the form of labeled example pages, or also by automatically labeling Web pages (such as WIEN), to build a training set.
In addition, it might exploits \textit{a priori} knowledge on the schema of the Web pages, for example taking into account previously learned page templates.
RoadRunner relies on the idea of working with two HTML pages at the same time in order to discover patterns analyzing similarities and differences between structure and content of each pair of pages.
Essentially, RoadRunner can extract relevant information from any Web site containing at least two Web pages with a similar structure.
Since usually Web pages are dynamically generated starting from template, and relevant data are positioned in the same (or in similar) areas of the page, RoadRunner is able to exploit this characteristic to identify relevant pieces of information, and, at the same time, taking into account small differences due to missing values or other mismatches.

The authors defined as \textit{class of pages} those Web sources characterized by a common generation script.
Then, the problem is reduced to extracting relevant data by generating wrappers for class of pages, starting from the inference of a common structure from the two-page-based comparison.
This system can handle missing and optional values and also structural differences, adapting very well to all kinds of data-intensive Web sources.
Another strength of RoadRunner is its high-quality open-source implementation (see: {\tt www.dia.uniroma3.it/db/roadRunner/}), that provides a high degree of reliability of the extraction system.

\paragraph*{Spatial reasoning} \label{sub:spatial-reasoning}
The paradigm of the Computer Vision has also inspired the field of Web Data Extraction systems.
In fact, a recent model of data extraction, called Visual Box Model, has been presented \cite{16,17}.
The Visual Box Model exploits visual cues to understand if, in the version of the Web page displayed on the screen, after the rendering of the Web browser, are present, for example, data in a tabular format.
The advantage of this strategy is that it is possible to acquire data not necessarily represented by means of the standard HTML \texttt{<table>} format.

The functioning of this technique is based on a X-Y cut OCR algorithm.
This algorithm is able, given the rendered version of a Web page, of generating a visual grid, where elements of the page are allocated according to their coordinates -- determined by visual cues.
Cuts are recursively applied  to the bitmap image representing the rendering of the Web page, and stored into an X-Y tree.
This tree is built so that ancestor nodes with leaves represent not-empty tables.
Some additional operations check whether extracted tables contain useful information.
This is done because -- although it is a deprecated practice -- many Web pages use tables for structural and graphical purposes, instead of for data representation scopes.

The Visual Box Model data extraction system is implemented by means of an internal rendering engine that produces a visualization of the Web page relying on Gecko ({\tt developer.mozilla.org/en/Gecko}), the same rendering engine used by the Mozilla Web browser.
By exploiting the CSS 2.0 box model, the algorithm is able to access the positional information of any given element.
This is achieved by a bridge between the rendering engine and the application, implemented by means of the  XPCOM library ({\tt developer.mozilla.org/it/XPCOM}).

\section{Web Data Extraction Systems}
\label{sec:webdataextraction}

In this section we get into details regarding the characteristics of existing Web Data Extraction systems.
We can generically define a Web Data Extraction system as a platform implementing \textit{a sequence of procedures (for example, Web wrappers) that extract information from Web sources} \cite{1}.
A large number of Web Data Extraction systems are available as commercial products even if an increasing number of free, open-source alternatives to commercial software is now entering into the market.
In most cases, the average end users of Web Data Extraction systems are firms or data analysts looking for relevant information from the Web.
An intermediate category of users consists of non-expert individuals that need to collect some Web content, often on a non-regular basis.
This category of users are often non-experienced and they look at simple but powerful Web Data Extraction software suites; among them we can cite DEiXTo\footnote{\url{http://deixto.com}}; DEiXTo is based on the W3C Document Object Model and enable users to easily create extraction rules pointing out the portion of data to scrape from a Web site.

In the next subsections we first illustrate the various phases characterizing a Web Data Extraction system (see Section \ref{sub:steps-web-data-extraction}) Subsequently, in Section \ref{sub:layer-cake}, we consider various factors influencing the design and implementation of Web Data Extraction system (e.g. the task of generating wrappers according to the easiness of use). We then illustrate the technological evolution of Web Data Extraction systems according to each factor under inquiry. To this purpose, we use ad hoc diagrams consisting of a group of layers ({\em layer cakes}) such that bottom (resp., top) layers correspond to the earliest (resp., latest) technological solutions.

\subsection{The main phases associated with a Web Data Extraction System}
\label{sub:steps-web-data-extraction}

In this section we describe the various phases associated with the procedure of extracting data from a Web site.

\paragraph*{Interaction with Web pages}
The first phase of a generic Web Data Extraction system is the \textit{Web interaction} \cite{26}: the Web Data Extraction system accesses a Web source and extracts information stored in it. Web sources usually coincide with Web pages, but some approaches consider also as RSS/Atom feeds \cite{31} and Microformats \cite{27}.

Some commercial systems, Lixto for first but also Kapow Mashup Server (described below), include a \textit{Graphical User Interface} for fully \textit{visual and interactive navigation} of HTML pages, integrated with data extraction tools.

The most advanced Web Data Extraction systems support the extraction of data from pages reached by \textit{deep Web navigation} \cite{37}: they simulate the activity of users clicking on DOM elements of pages, through macros or, more simply, by filling HTML forms.

These systems also support the extraction of information from \textit{dynamically generated Web pages}, usually built at run-time as a consequence of the user request, filling a template page with data from some database.
The other kind of pages are commonly called \textit{static Web pages}, because of their static content.

OxPath \cite{Oxpath}, which is part of the DIADEM project \cite{diadem} is a declarative formalism that extends XPath to support deep Web navigation and data extraction from interactive Web sites. It adds new location steps to simulate user actions, a new axis to select dynamically computed CSS attributes, a convenient field function to identify visible fields only, page iteration with a Kleene-star extension for repeated navigations, and new predicates for marking expressions for extraction identification.

\paragraph*{Generation of a wrapper}
A Web Data Extraction system must implement the support for wrapper generation and wrapper execution.

Another definition of Web Data Extraction system was provided by Baumgartner et al. \cite{8}. They defined a Web Data Extraction system as ``a software extracting, \textit{automatically} and \textit{repeatedly}, data from Web pages with changing contents, and that \textit{delivers extracted data} to a database or some other application''.
This is the definition that better fits the modern view of the problem of the Web Data Extraction as it introduces three important aspects:

\begin{itemize}
    \item Automation and scheduling
    \item Data transformation, and the
    \item Use of the extracted data
\end{itemize}

In the following we shall discuss each of these aspects into detail.

{\em Automation and Extraction}. Automating the access to Web pages as well as the localization of their elements is one of the most important features included in last Web Data Extraction systems \cite{36}. 

Among the most important automation features we cite the possibility to \textit{simulate the click stream of the user}, \textit{filling forms} and \textit{selecting menus and buttons}, the support for AJAX technology \cite{35} to handle the asynchronous updating of the page and the ability of scheduling Web data extraction procedures on a periodical basis.

{\em Data transformation}. Information could be wrapped from multiple sources, which means using different wrappers and also, probably, obtaining different structures of extracted data. The steps between extraction and delivering are called \textit{data transformation}: during these phases, such as \textit{data cleaning} \cite{41} and \textit{conflict resolution} \cite{42}, users reach the target to obtain homogeneous information under a unique resulting structure. The most powerful Web Data Extraction systems provide tools to perform \textit{automatic schema matching} from multiple wrappers \cite{40}, then packaging data into a desired format (e.g., a database, XML, etc.) to make it possible to \textit{query data}, \textit{normalize structure} and \textit{de-duplicate tuples}.

{\em Use of extracted data}. When the extraction task is complete, and acquired data are packaged in the needed format, this information is ready to be used; the last step is to deliver the package, now represented by structured data, to a managing system (e.g., a native XML DBMS, a RDBMS, a data warehouse, a CMS, etc.). In addition to all the specific fields of application covered later in this work, acquired data  can be also generically used for analytical or statistical purposes \cite{44} or simply to republish them under a structured format.

\subsection{Layer cake comparisons}
\label{sub:layer-cake}

In this section, we summarize the capability stacks of Web data extraction systems from our understanding, including aspects of wrapper generation, data extraction capabilities, and wrapper usage. In particular, we introduce some specific aspects and illustrate the technological evolution of Web Data Extraction systems according to each aspect. We use ad hoc diagrams structured as a group of layers ({\em layer cakes}). In these diagrams, bottom (resp., top) layers correspond to the earliest (resp., latest) technological solutions.

\begin{figure}[ht] \centering
	\begin{minipage}{.49\columnwidth}
		\includegraphics[width=\columnwidth, trim=0 250 0 50]{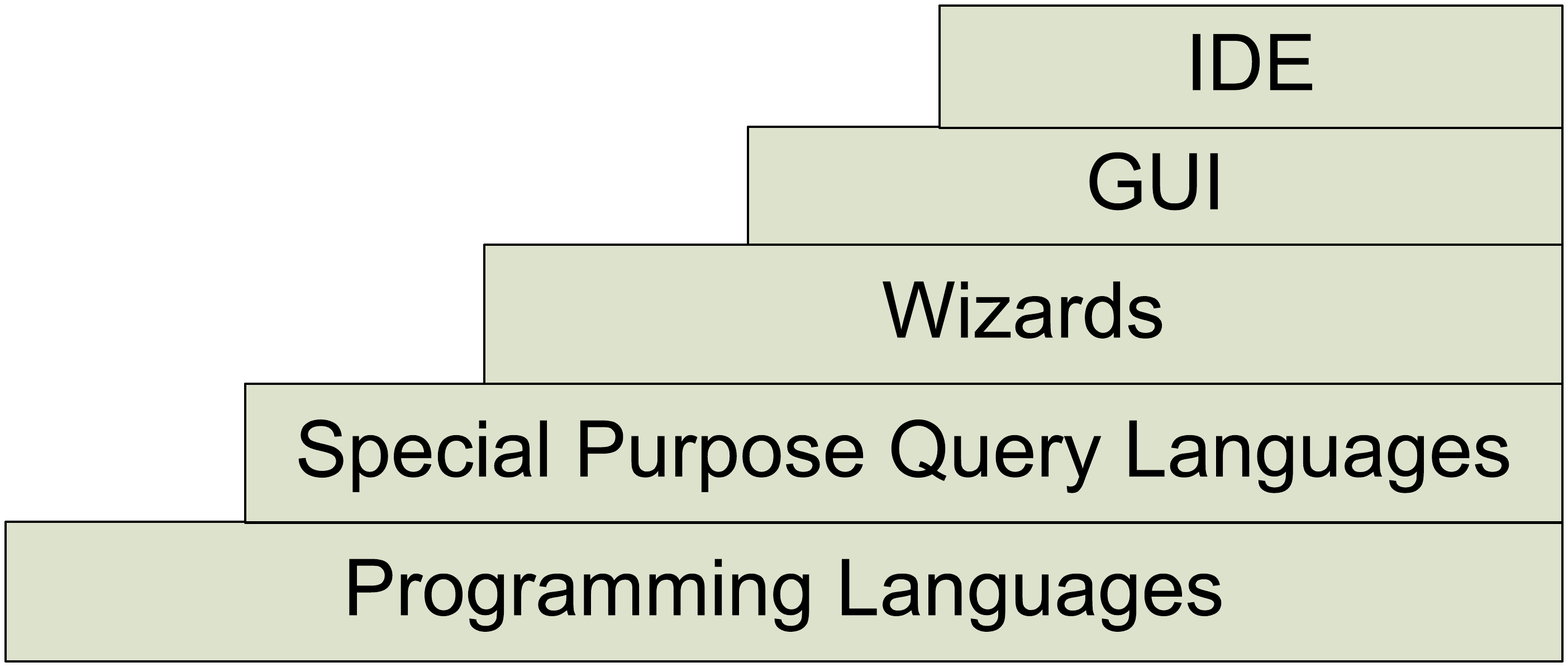}
		\caption{Wrapper Generation: Ease of Use}
		\label{fig:cake1}
	\end{minipage}
	\begin{minipage}{.49\columnwidth}
		\includegraphics[width=\columnwidth, trim=0 250 0 40]{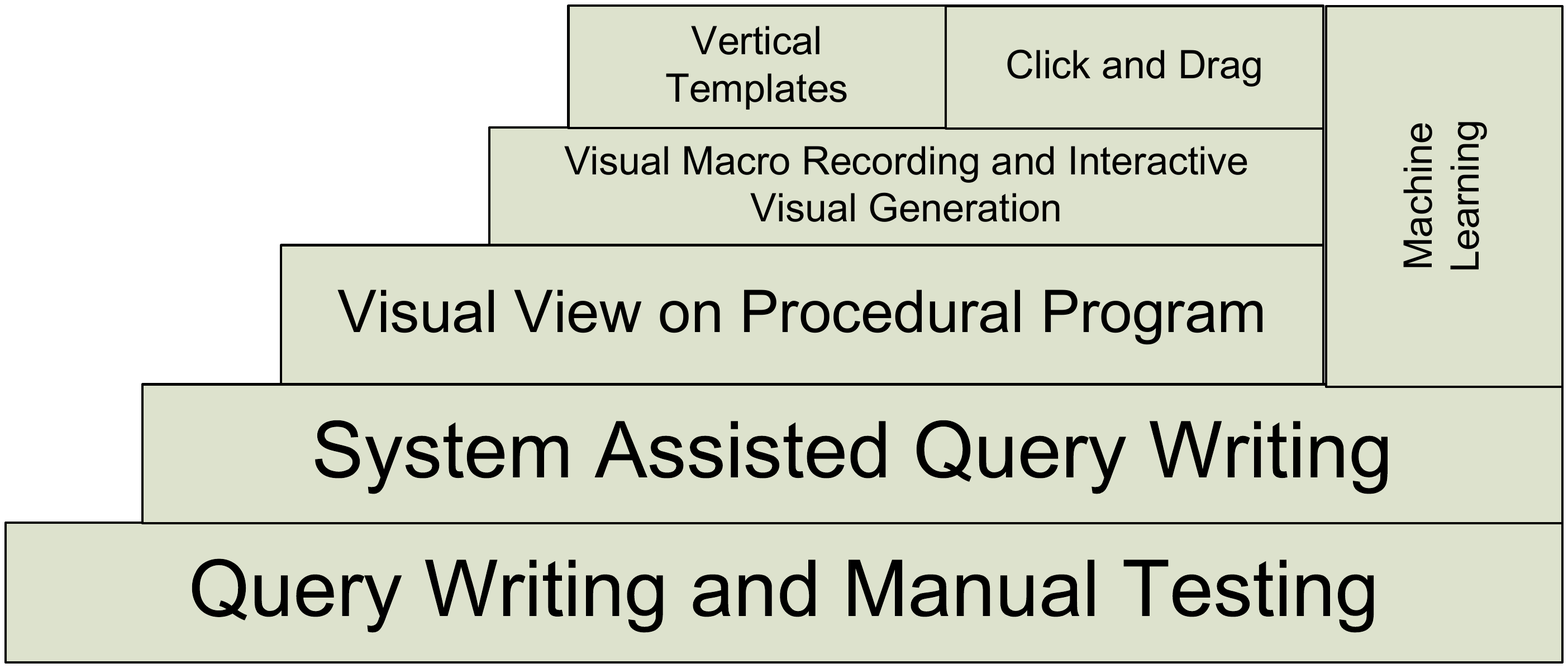}
		\caption{Wrapper Creation Paradigma}
		\label{fig:cake2}
	\end{minipage}
\end{figure}

\paragraph*{Wrapper Generation: Ease of Use (Figure \ref{fig:cake1})}
The first approaches to consuming facts from the Web were implemented by means of general purpose languages. Over time libraries (e.g.\ Ruby Mechanize) and special-purpose query languages evolved on top of this principle (e.g., Jedi \cite{Jedi} and Florid \cite{Florid}). Wizards that simplify the way to specify queries are the next logical level and for instance have been used in W4F \cite{11} and XWrap \cite{LPHBW}. Advanced Web Data Extraction systems offer GUIs for configuration, either client-based (e.g.\ Lapis), Web-based (e.g.\ Dapper and Needlebase) or as browser extensions (e.g.\ iOpus and Chickenfoot). Commercial frameworks offer a full IDE (Integrated Development Environment) with functionalities described in the previous section (e.g.\ Denodo, Kapowtech, Lixto and Mozenda).

\paragraph*{Wrapper Generation: Creation Paradigma (Figure \ref{fig:cake2})}
From the perspective of how the system supports the wrapper designer to create robust extraction programs, the simplest approach is to manually specify queries and test them against sample sites individually.
Advanced editors offer highlighting of query keywords and operators and assist query writing with auto-completion and similar usability enhancements (e.g., Screen-Scraper). In case of procedural languages, debugging means and visual assistance for constructs such as loops are further means to guided wrapper generation. Regarding the definition of deep Web navigations, such as form fillouts, a number of tools offer VCR-style recording the human browsing and replaying the recorded navigation sequence (e.g., Chickenfoot, iOpus, Lixto). Visual and interactive facilities are offered by systems to simplify fact extraction. Users mark an example instance, and the system identifies the selected element in a robust way, and possibly generalize to match further similar elements (e.g., to find all book titles). Such means are often equipped with Machine Learning methods, in which the user can select a multitude of positive and negative examples, and the system generates a grammar to identify the Web objects under consideration, often in an iterative fashion (e.g.\ Wien, Dapper, Needlebase). Click and drag and drop features further simplify the interactive and visual mechanisms. Finally, some systems offer vertical templates to easily create wrappers for particular domains, for example for extracting hotel data or news item, using Natural Language Processing techniques and domain knowledge, or for extracting data from typical Web layouts, such as table structures or overview pages with next links.

\begin{figure}[ht] \centering
	\begin{minipage}{.49\columnwidth}
		\includegraphics[width=\columnwidth, trim=0 300 0 50]{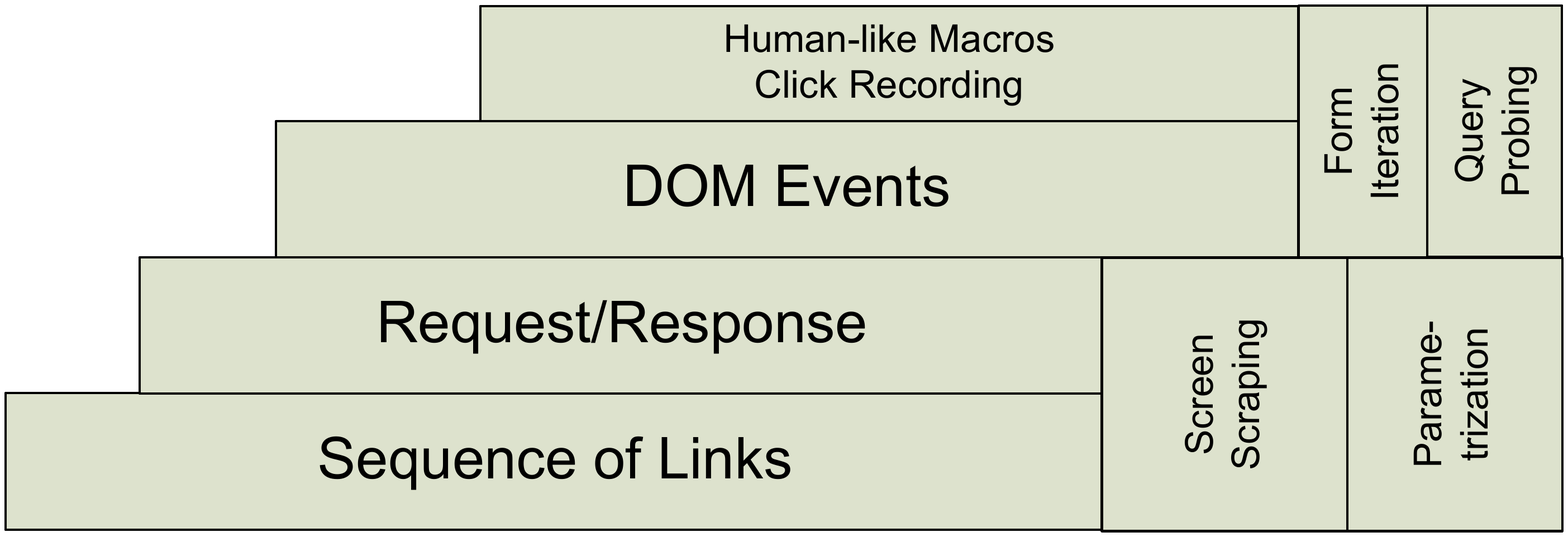}
		\caption{Deep Web Navigation Capabilities}
		\label{fig:cake3}
	\end{minipage}
	\begin{minipage}{.49\columnwidth}
		\includegraphics[width=\columnwidth, trim=0 275 0 50]{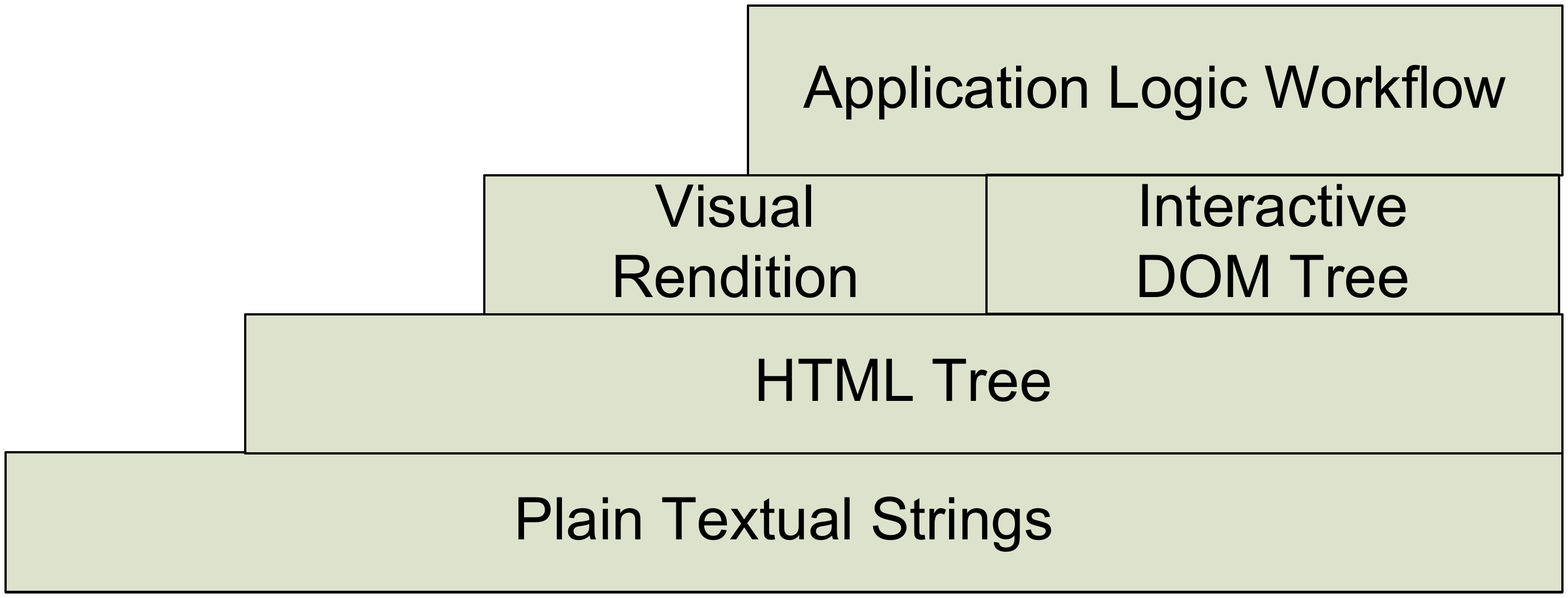}
		\caption{Web Data Extraction Capabilities}
		\label{fig:cake4}
	\end{minipage}
\end{figure}

\paragraph*{Deep Web Navigation Capabilities (Figure \ref{fig:cake3})}
Before the advent of Web 2.0 techniques, dynamic HTML and AJAX it was usually sufficient to consider the Web as a collection of linked pages. In such cases, form filling can be simulated by tracking the requests and responses from the Web Server, and replaying the sequence of requests (sometimes populating a session id dynamically, extracted from a previous page of the sequence). Alternatively, early Web Data extraction systems have been influenced by screen scraping technologies like they were being used for automating 3270 applications or like used for automating native applications, usually relying heavily on coordinates. Understanding and replaying DOM Events on Web objects is the next logical level in this capability stack. Advanced systems even go a step further, especially when embedding a full browser: the click on an element is recorded in a robust way and during replay the browser is informed to do a visual click on such an element, handing the DOM handling over to the browser and making sure that the Web page is consumed exactly in the way the human user consumes it. Orthogonal to such features are capabilities to parametrize deep Web sequences, and to use query probing techniques to automate deep Web navigation to unknown forms.

\paragraph*{Web Data Extraction Capabilities (Figure \ref{fig:cake4})}
Over time, various approaches to modeling a Web page have been discussed. The simplest way is to work on the stream received from the Web server, for example using regular expressions.
In some cases, this is sufficient and even the preferred approach in large-scale scenarios due to avoiding to build up a complex and unperformant model. On the other hand, in complex Web 2.0 pages or in pages that are not well-formed, it can be extremely cumbersome to work on the textual level only. Moreover, such wrappers are not very maintainable and break frequently. The most common approach is to work on the DOM tree or some other kind of tree structure.
This approach has been followed both by the academic community and by commercial approaches. In the academic communities studying expressiveness of language over trees and tree automata are interesting, and from a commercial point of view, it is convenient and robust to use languages such as XPath for identifying Web objects. Usually, not only the elements of a DOM tree are considered, but also the events, enabling to specify data extraction and navigation steps with the same approach.
In the Web of today, however, very often the DOM tree does not really capture the essential structure of a Web page as presented to the human user in a browser.
A human perceives something as table structure, whereas the DOM tree contains a list of {\tt div} elements with absolute positioning. Moreover, binary objects embedded into Web pages such as Flash pose further challenges and are not covered with a tree structure approach. Hence, screen-scraping made back its way into novel Web Data extraction frameworks, using methods from document understanding and spatial reasoning
such as the approaches of the TamCrow project  \cite{kruepl}, of the ABBA project \cite{67} spatial XPath extensions \cite{SXpath} and rendition-based extensions in RoadRunner to detect labels \cite{roadrunnervisuallabels}.

\begin{figure}[ht] \centering
	\begin{minipage}{.49\columnwidth}
		\includegraphics[width=\columnwidth, trim=0 275 0 50]{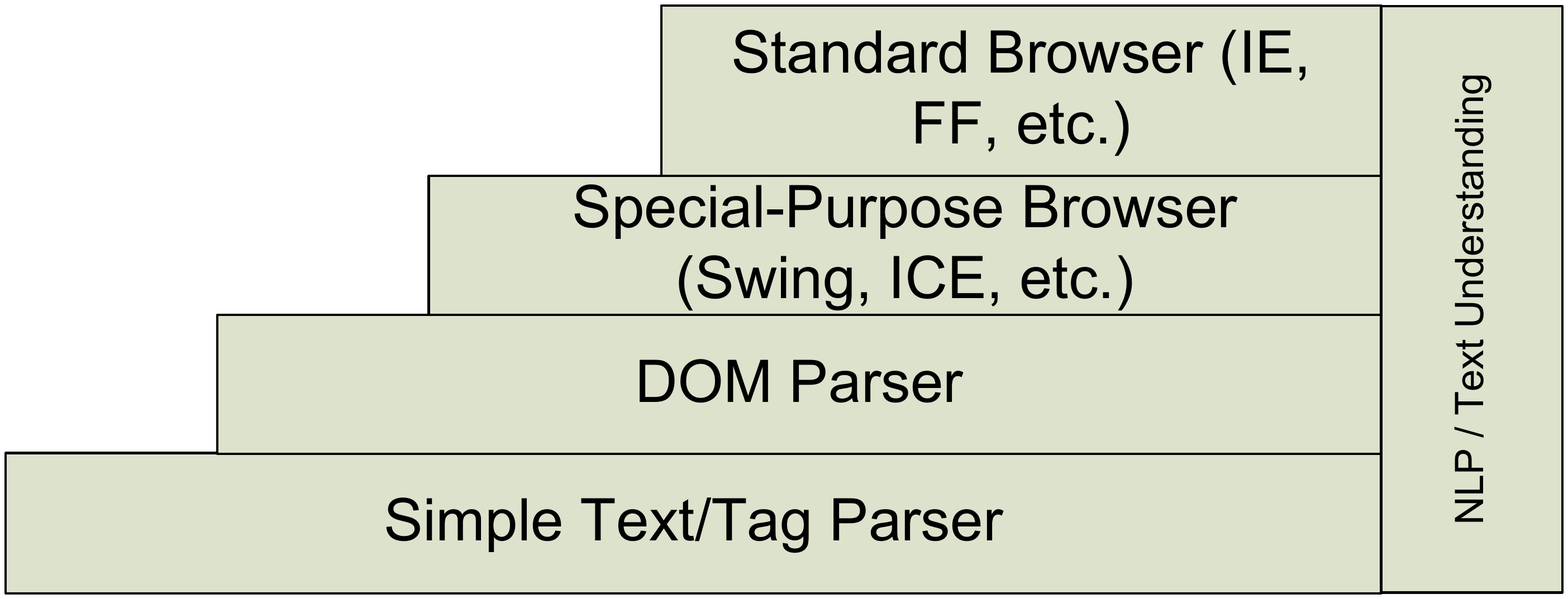}
		\caption{Parser and Browser Embedding}
		\label{fig:cake5}
	\end{minipage}
	\begin{minipage}{.49\columnwidth}
		\includegraphics[width=\columnwidth, trim=0 275 0 50]{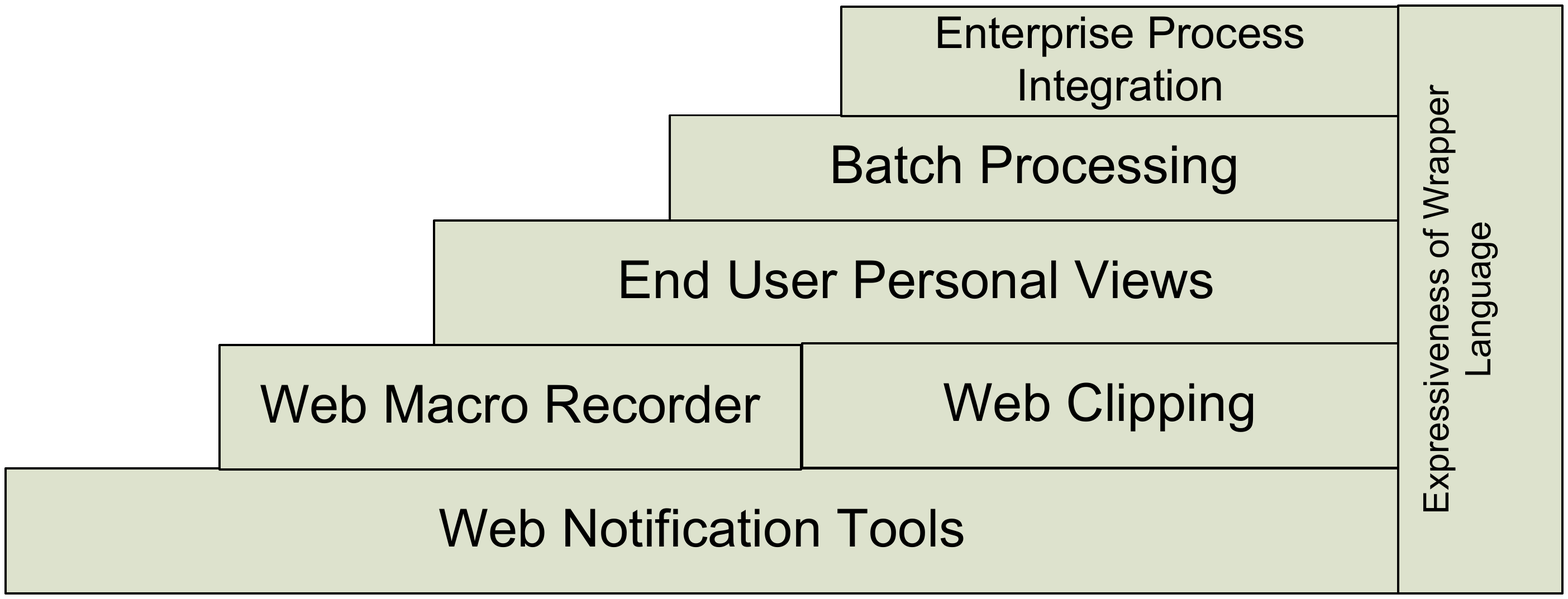}
		\caption{Complexity of Supported Operations}
		\label{fig:cake6}
	\end{minipage}
\end{figure}

\paragraph*{Parser and Browser Embedding (Figure \ref{fig:cake5})}
This capability stack is closely related to the Deep Web capabilities but focussing on the technical realization of parsing and browser embedding. Simple approaches create their own parser to identify relevant HTML tags, more sophisticated approaches use DOM libraries without an associated browser view. Due to the fact that many Web Data Extraction frameworks are implemented in Java, special-purpose browsers such as the Java Swing browser and the ICE browser are
and have been used. The most powerful approaches are the ones that embed a standard browser such as IE, Firefox or WebKit-based browsers. In case of Java implementations, interfaces such as Java-XPCOM bridge or libraries such as JRex are used to embed the Mozilla browser. Embedding a full browser not only gives access to the DOM model, but additionally to other models useful for data extraction, including the CSS Box model. Some tools go a different direction and instead of embedding a browser, they are implemented as browser extension, imposing some restrictions and inconveniences. Orthogonal to the browser stack are the capabilities to extend extraction functionalities to unstructured text parsing exploiting Natural Language Processing techniques.

\paragraph*{Complexity of Supported Operations (Figure \ref{fig:cake6})}
Simple data extraction tools offer Web notification, e.g.if a particular word is mentioned. Web macro recorders allow users to create ``deep'' bookmarks, and Web clipping frameworks clip fragments of a Web page to the desktop. Personalization of a Web page (e.g., some tools offer to alter the CSS styles to make frequently used links more prominent on a page) is a further level in the layer cake. Batch processing frameworks offer functionalities to replay ample extraction tasks (e.g., running through many different values in form fillout). Advanced systems go a step further and use sophisticated extraction plans and anonymization techniques that ensure to stay under the radar and not harm target portals with too many requests at once, and deliver the data into further applications such as market intelligence platforms. The expressiveness of a wrapper language contributes as well to the complexity of supported operations.

\section{Applications}
\label{sec:applications}

The aim of the second part of this paper is to survey and analyze a large number of applications that are strictly interconnected with Web Data Extraction tasks. To the best of our knowledge, this is the first attempt to classify applications based on Web Data Extraction techniques even if they have been originally designed to operate in specific domain and, in some cases, they can appear as unrelated.

The spectrum of applications possibly benefiting from Web Data Extraction techniques is quite large.
We decided to focus on two main application scenarios, namely the {\em enterprise domain} and {\em the Social Web domain} because, in our opinion, these scenarios can take advantage the most from Web Data Extraction technologies.
As for the enterprise domain, Web Data Extraction software can significantly increase the efficiency of business processes by freeing up resources like time and manpower. Due to the automatic collection of data from the Web, companies have at their disposal large amount of data in a relatively short time frame and these data assist firm analysts and manager to plan or revise business strategies.
As for the Social Web domain, extracting data from Social Web sites/Social Media is a relatively new concept that has acquired a big relevance due to the explosive growth of platforms like Facebook, Twitter or Instagram. Crawling social media and online social networks data has become the flagship Social application of Web Data Extraction \cite{Catanese2011,ferrara2013clustering}. This yield the unprecedented opportunity for scientists from different domains (e.g., Sociology, Political Science, Anthropology, etc.) to analyze and understand the dynamics of human behaviors at a planetary scale and in a real time-fashion \cite{ferrara2012large,conover2013geospatial,demeo2013analyzing,ferrara2013traveling}.
The classification scheme above, however, has not to be intended as hard because there are some applications which have been originally designed to work in the enterprise domain but they have subsequently (and successfully) re-used in the context of the Social Web.
For instance, companies can extract data from Social Web platforms to pursuit tasks like {\em reputation management} (i.e., a company is informed of who is talking about the firm itself and its products) and {\em trend management} (i.e., a company can identify what are the topics catching the interest of its customers). In general, user generated content available on Social Web platforms are perhaps the data that firms are interested the most: data such as product or service reviews, feedback and complaints are a key ingredient to monitor the impact of a brand on the market and the sentiments toward that brand and to detect how a brand is perceived in opposition to its commercial competitors.
On the other hand, data produced by users who discuss on the same product/service/brand have also a social dimension because they reflect social interactions: a user in a given community (e.g., a discussion forum of a company) can pose questions to other members on the same community and they can rank the advice they receive on the basis of the social relationship like trust or friendship binding the user who pose the question with the users who answered it.

In addition to classifying applications into the enterprise and Social Web categories, we can introduce other criteria which rely on the fact that data are extracted from a single or multiple sources, data can be of the same format or, finally, are associated with the same domain. In particular, we introduce the following criteria:

\paragraph*{Single Source vs. Multiple Sources}
Web Data Extraction techniques can be applied on data residing on a single platform or, vice versa, they can collect data located in different platforms.

    As an example, in the context of Enterprise applications, some applications fetch data from a single platform: a relevant example is provided by applications to manage customer care activities. As previously pointed out, these application crawls a corpus of text documents produced within a single platform. By contrast, in some cases, the application can benefit of data scattered across multiple system. In the context of Enterprise applications, a notable example of application gathering data from multiple sources is given by {\em Comparison Shopping}: in such a case, a Web Data Extraction technique is able to collect data associated with a target product from a multitude of e-commerce platforms and to compare, for instance, the price and the shipping details of that product in each of these platform.

    The classification above is particularly interesting in the field of Social Web applications: in fact we discuss separately applications designed to crawl and collect data from a single platform (see Section \ref{sub:socialnetwork}) from applications running on multiple Social Web platforms.

\paragraph*{Homogeneity in Format vs. Heterogeneity in Format}
The second classification criterium we introduce answers the following question: does the application collect data of the same format or, vice versa, can it collect data of different formats? On the basis of such a criterium we can classify existing applications in {\em homogeneous-format} and {\em heterogeneous-format}.

    As for enterprise applications, a nice example of homogeneous-format applications is given by {\em Opinion Sharing Applications}. This term identifies applications devoted to collect the opinions a user expresses about a given service/product. These opinions are typically represented by short textual reviews or blog post. In such a case, the type of data extracted by a Web Data Extraction technique is a string. In some cases, users are allowed to provide scores and the format of extracted data is a discrete value, generally ranging in an interval like [0,5]. In case of Social Web applications, many approaches (like \cite{Gjoka2010,Catanese2011,Mislove2007,Ye2010}) are devoted to extract friendship relationships from a Social Networking Websites. In such a case, extracted data are of the same format that can be interpreted as a triplet of the form $\langle u_x, u_y, f \rangle$ being $u_x$ and $u_y$ the user identifiers of two Social Network members; the value $f$ can be a boolean (e.g., it can be {\tt true} if $u_x$ and $u_y$ are friends, {\tt false} otherwise) or a numerical value (e.g., it can be set equal to the number of messages $u_x$ sent $u_y$).

    Some applications are designed to collect data of different type. A relevant example in the field of enterprise applications is provided by those applications designed to manage Business Intelligence tasks. These application are able to collect data of different type like (numerical data or textual ones) as well as to manage both structured data (e.g., data coming from relational tables of a DBMS) as well as unstructured ones (e.g., text snippets present in HTML pages). Relevant examples of applications capable of extracting data of different type are also present in the context of Social Web applications: for instance, \cite{Kwak10www} performed in 2009 a crawl of the whole Twitter platform which produced textual data (e.g., the tweets and re-tweets produced by users) as well as data indicating the different types of connections among users (``following'', ``reply to'' and ``mention'') \cite{romero2010directed}.

\paragraph*{Single Purpose or Multiple Purpose}
In some cases, the goal of a Web Data Extraction tool is to extract data describing a specific facet of a particular social phenomenon or business process. In such a case, an application is catalogued as {\em unique purpose}.
    Other applications, instead, aim at collecting data of different nature which, if properly linked, are relevant to better understand and interpret a particular phenomenon. After linking data, novel and sophisticated applications running on them can be defined. Applications belonging to this category will be called {\em multi purpose}.
    As an example of single purpose applications, we focus on applications devoted to collect bibliographic data as well as citations among papers. To collect bibliographical information, Web Data Extraction techniques are required to query several databases containing scientific publications (like ISI Web of Science, SCOPUS, PubMed and so on) but the goal of the application, however, is to collect all the citations associated with a particular paper or to a particular author because this information is useful to assess the impact of an author or a scientific publication in a scientific discipline.
    In the field of Social Web applications, an example of single purpose application is given by applications devoted at collecting data about human activities in different Social Web platforms: for instance, the tags contributed by a user in different systems \cite{Szomszor08}. This is relevant to understand if the language of a user is uniform across various platforms or if the platform features impact on the user formation of the vocabulary.

    We can cite some relevant examples of multi purpose applications both at the enterprise and at the Social Web level. For instance, in the class of enterprise applications, some applications are able to collect data produced by different Web Services and combine these data to produce more advanced applications. For instance, in the travel industry, one can think of applications collecting data on flights and hotels and combine these data to produce holiday packages. In the field of Social Web applications, several authors observed that the merging of different type of data describing different type of human activities produces a more detailed knowledge of human needs and preferences. For instance, in \cite{schifanella2010folks} the authors conducted extensive experiments on data samples extracted from Last.Fm and Flickr. They showed that a  strong correlations exist between user social activities (e.g., the number of friends of a user or the number of groups she joined) and the tagging activity of the same user. A nice result provided in \cite{schifanella2010folks} was that user contributed tags are a useful indicator to predict friendship relationships. Other studies entwining users geographical locations with the content they posted are provided in \cite{crandall2009mapping} and \cite{kinsella2011m}.

\subsection{Enterprise Applications}
\label{sec:enterprise}

In this section we describe the main features of software applications and procedures related with Web Data Extraction with a direct or subsequent \textit{commercial scope}.

\subsubsection{Context-aware advertising}

Context-aware advertising techniques aim at presenting to the end user of a Web site commercial \textit{thematized advertisements} together with the content of the Web page the user is reading. The ultimate goal is to increase the value that a Web page can have for its visitors and, ultimately, to raise the level of interest in the ad.

First efforts to implement context-aware advertising were made by Applied Semantic, Inc. ({\tt www.appliedsemantics.com}); subsequently Google bought their ``AdSense'' advertising solution.
The implementation of context-aware advertisements requires to \textit{analyze the semantic content} of the page, \textit{extract relevant information}, both in the structure and in the data, and then contextualize the ads content and placement in the same page.
Contextual advertising, compared to the old concept of Web advertising, represents a \textit{intelligent approach} to providing useful information to the user, statistically more interested in thematized ads, and a better source of income for advertisers.

\subsubsection{Customer care}

Usually medium- and big-sized companies, with customers support, handle a large amount of unstructured information available as text documents. Relevant examples are emails, support forum discussions, documentation, shipment address information, credit card transfer reports, phone conversation transcripts, an so on. The ability of analyzing these documents and extracting the main concepts associated with them provides several concrete advantages. First of all, documents can be classified in a more effective fashion and this makes their retrieval easier. In addition, once the concepts present in a collection of documents have been extracted, it is possible to identify relevant associations between documents on the basis of the concepts they share. Ultimately, this enables to perform sophisticated data analysis targeted at discovering trends or, in case of forms, hidden associations among the products/services offered by a brand.
In this context, Web Data Extraction techniques play a key role because they are required to quickly process large collections of textual documents and derive the information located in these documents. The retrieved data are finally processed by means of algorithms generally coming from the area of Natural Language Processing.

\subsubsection{Database building}

In the Web marketing sector, Web Data Extraction techniques can be employed to gather data referring to a given domain.
These data may have a twofold effect: {\em (i)} a design, through reverse engineering analysis, can design and implement a DBMS representing that data; {\em (ii)} the DBMS can be automatically populated by using data provided by the Web Data Extraction system.
Activities {\em (i)} and {\em (ii)} are also called {\em Database Building}.
Fields of application of Database Building are countless: financial companies could be interested in extracting financial data from the Web and storing them in their DBMSs.
Extraction tasks are often scheduled so that to be executed automatically and periodically.
Also the real estate market is very florid: acquiring data from multiple Web sources is an important task for a real estate company, for comparison, pricing, co-offering, etc.
Companies selling products or services probably want to \textit{compare their pricing with other competitors}: products pricing data extraction is an interesting application of Web Data Extraction systems. Finally we can list other related tasks, obviously involved in the Web Data Extraction: duplicating an on-line database, extracting dating sites information, capturing auction  information and prices from on-line auction sites, acquiring job postings from job sites, comparing betting information and prices, etc.

\subsubsection{Software Engineering}

Extracting data from Web sites became interesting also for Software Engineering: for instance, Rich Internet Applications (RIAs) are rapidly emerging as one of the most innovative and advanced kind of application on the Web. RIAs are Web applications featuring a high degree of interaction and usability, inherited from the similarity to desktop applications. Amalfitano et al. \cite{15} have developed a reverse engineering approach to abstract Finite States Machines representing the \textit{client-side behavior} offered by RIAs.

\subsubsection{Business Intelligence and Competitive Intelligence}

Baumgartner et al. \cite{45,62,63} deeply analyzed how to apply Web Data Extraction techniques and tools to improve the process of
acquiring market information. A solid layer of knowledge is
fundamental to optimize the decision-making activities and a large amount of public information could be retrieved on the Web.
They illustrate how to acquire these unstructured and semi-structured information.
In particular, using the Lixto Suite to access, extract, clean and deliver data, it is possible to gather, transform and obtain information useful to business purposes. It is also possible to integrate these data with other common platforms for Business Intelligence, like SAP ({\tt www.sap.com}) or Microsoft Analysis Services \cite{48}.

Wider, the process of gathering and analyzing information about products, customers, competitors with the goal of helping the managers of a company in decisional processes is commonly called {\em Competitive Intelligence}, and is strictly related to data mining \cite{51}. Zanasi \cite{47} was the first to introduce the possibility of acquiring these data, through data mining processes, on public domain information. Chen et al. \cite{46} developed a platform, that works more like a spider than a Web Data Extraction system, which represents a useful tool to support Competitive Intelligence operations.
In Business Intelligence scenarios, we ask Web Data Extraction techniques to satisfy two main requirements: scalability and efficient planning strategies because we need to extract as much data as possible with the smallest amount of resources in time and space.

\subsubsection{Web process integration and channel management}

In the Web of today data is often available via APIs (e.g., refer to: {\tt www.programmableweb.com}). Nevertheless, the larger amount of data is primarily available in semi-structured formats such as HTML. To use Web data in Enterprise Applications and service-oriented architectures, it is essential to provide means for automatically turning Web Applications and Web sites into Web Services, allowing a structured and unified access to heterogeneous sources. This includes to understand the logic of the Web application, to fill out form values, and to grab relevant data.

In a number of business areas, Web applications are predominant among business partners for communication and business processes. Various types of processes are carried out on Web portals, covering activities such as purchase, sales, or quality management, by manually interacting with Web sites. Typical vertical examples, where Web Data Extraction proves useful include channel management in the travel industry (like automating the regular offerings of rooms on hotel portals with bi-directional Web connectors), re-packaging complex Web transactions to Web services and consequently to other devices, as well as automating communication of automotive suppliers with automotive companies.
Tools for wrapper generation pave the way for \emph{Web Process Integration} and enable the \emph{Web of Services}, i.e., the seamless integration of Web applications into a corporate infrastructure or service oriented landscape by generating Web services from given Web sites \cite{66}. Web process integration can be understood as front-end and ``outside-in'' integration: integrate cooperative and non-cooperative sources without the need for information provider to change their backend.
Additional requirements in such scenarios include to support a large number of users and real-time parametrized Web queries and support of complex Web transactions.

\subsubsection{Functional Web application testing}

Testing and Quality Management are essential parts of the life-cycle of software. Facets of testing are manifold, including functional tests, stress/load tests, integration tests, and testing against specifications, to name a few.
Usually, the strategy is to automate a large percentage of functional tests and execute test runs as part of nightly builds as regression tests.
Such tests occur at various levels, for example testing the system functionality as a blackbox via APIs, or testing the system at the GUI level simulating either the user's steps or creating a model of possible application states.

In today's world of Software-as-a-Service platforms and Web oriented architectures, Web application testing plays an important role. One aspect is simulating the user's path through the application logic.
Robust identification criteria can be created by taking advantage of the tree structure or visual structure of the page. Typical actions in such test scripts include to set/get values of form fields, picking dates, checkpoints to compare values, and following different branches depending on the given page. Due to automation, every step can be parametrized and a test script executed in variations.
The requirements for tools in the area of Web application testing are to deal well with AJAX/dynamic HTML, to create robust test scripts, to efficiently maintain test scripts, to execute test runs and create meaningful reports, and, unlike other application areas, the support of multiple state-of-the-art browsers in various versions is an absolute must. One widely used open source tool for Web application testing is \emph{Selenium} ({\tt seleniumhq.org}).

\subsubsection{Comparison shopping}

One of the most appreciated services in e-commerce area is the \textit{comparison shopping}, i.e.,  the capability of comparing products or services; various type of comparisons are allowed going from simple prices comparison to features comparison, technical sheets comparison, user experiences comparison, etc.
These services heavily rely on Web Data Extraction, using Web sites as sources for data mining and a custom internal engine to make possible the comparison of similar items.
Many Web stores today also offer \textit{personalization forms} that make the extraction tasks more difficult: for this reason many last-generation commercial Web Data Extraction systems  (e.g., Lixto, Kapow Mashup Server, UnitMiner, Bget) provide support for deep navigation and dynamic content pages.

\subsubsection{Mashup scenarios}

Today, leading software vendors provide mashup platforms (such as Yahoo! Pipes or Lotus Mashups) and establish mashup communication standards such as EMML (see: {\tt www.openmashup.org}). A mashup is a Web site or a Web application that combines a number of Web sites into an integrated view. Usually, the content is taken via APIs, embedding RSS or Atom Feeds in a REST-like way. With wrapper technology, one can leverage legacy Web applications to light-weight APIs such as REST that can be integrated in mashups in the same fashion. Web Mashup Solutions no longer need to rely on APIs offered by the providers of sites, but can extend the scope to the whole Web. In particular, the deep Web gets accessible by encapsulating complex form queries and application logic steps into the methods of a Web Service.
End users are put in charge of creating their own views of the Web and embed data into other applications (``consumers as producers''), usually in a light-weight way. This results in ``situational applications'', possibly unreliable and unsecure applications, that however help to solve an urgent problem immediately. In Mashup scenarios, one important requirement of Web Data Extraction tools is the ease of use for non-technical content managers, to give them the possibility to create new Web connectors without help of IT experts.

\subsubsection{Opinion mining}

Related to comparison shopping, the opinion sharing represents its evolution: users want to \textit{express opinions} on \textit{products}, \textit{experiences}, \textit{services} they enjoyed, etc.
The most common form of opinion sharing is represented by blogs, containing articles, reviews, comments, tags, polls, charts, etc.
As an example, we can cite MOpiS (Multiple Opinion Summarizer) \cite{kokkoras2008mopis}, an algorithm that generates summaries of reviews associated with commercial products by taking into account both the review content and some metadata (e.g., the usefulness of a review, the technical expertise of the reviewer and so on) that are usually present together with review text.
All this information usually lacks of structure, so their extraction is a big problem, also for current systems, because of the billions of Web sources now available.
Sometimes model-based tools fit good, taking advantage of common templates (e.g., Wordpress ({\tt www.wordpress.org}), Blogger ({\tt www.blogger.com}), etc.), other times Natural Language Processing techniques fit better. Kushal et al. \cite{61} approached the problem of opinion extraction and subsequent semantic classification of reviews of products.

Another form of opinion sharing in semi-structured platforms is represented by Web portals that let users to write unmoderated opinions on various topics.

\subsubsection{Citation databases}

Citation databases building is an intensive Web Data Extraction field of application: CiteSeer ({\tt citeseer.ist.psu.edu}), Google Scholar and DBLP ({\tt www.informatik.uni-trier.de/ley/db}), amongst others, are brilliant examples of applying Web Data Extraction to approach and solve the problem of collect digital publications, extract relevant data, i.e.,  \textit{references and citations}, and build structured databases, where users can perform searches, comparisons, count of citations, cross-references, etc.

Several challenges are related to this context of application: for example, the corpus of scientific publications could rapidly vary over time and, to keep the database updated it could be necessary to repeatedly apply a Web Data Extraction process {\em from scratch} on the same Web sources.
Such an operation, however, can be excessively time-consuming.
An attempt to address this challenge has been done in \cite{chen2008efficient}: in that paper, the authors suggest an {\em incremental solution} which requires to identify portions of information shared by consecutive snapshots and to reuse the information extracted from a snapshot to the subsequent one.

\subsubsection{Web accessibility}

Techniques for automatic data extraction and document understanding are extremely helpful in making Web pages more accessible to blind and partial-sighted users.

Today's solution approaches are inefficient to overcome the problem. The first approach, screen-reader usage, is optimized for native client interfaces, and not well equipped to deal with the presentation, content and interactions in Web 2.0 - such as understanding the reading order, telling the user what a date picker means, or jumping from one forum post to the next. The second approach, the Web Accessibility Initiative, is no doubt absolutely necessary and has defined valuable concepts, such as ARIA roles assignable to GUI elements. However, due to the additional investment at best such guidelines are applied in governmental sites.

Approaches such as ABBA \cite{67} overcome these limitations. In the ABBA approach, a Web page is transformed into a formal multi-axial semantic model; the different axes offer means to reason on and serialize the document by topological, layout, functional, content, genre and saliency properties. A blind person can navigate along and jump between these axes to skip to the relevant parts of a page. E.g., the presentational axis contains transformed visual cues, allowing the user to list information in the order of visual saliency.

\subsubsection{Main content extraction}

Typical Web pages, for example news articles, contain, additionally to the main content, navigation menus, advertisements and templates. In some cases, such as when archiving a news article Web page for later offline reading it is convenient to get rid of such irrelevant fragments.
To extract the main content only, one needs to apply techniques to distinguish the relevant content from the irrelevant one. Approaches range from complex visual Web page analysis to approaches leaning on text density or link density analysis. An approach for boilerplate detection using shallow text features was introduced in \cite{kohlschutter2010boilerplate}\footnote{The associated software can be found at \url{http://code.google.com/p/boilerpipe/}}.
Additionally, tools/apps such as InstaPaper or the Readability Library -- The reader is referred to the PHP port {\tt www.keyvan.net/2010/08/php-readability} -- use main content extraction to store the relevant fragment and text from a Web page that resembles the article for later reading.

\subsubsection{Web (experience) archiving}

Digital preservation and Web data curation are the goals of the discipline of Web Archiving. On the one hand, this means to access information no longer available in the live Web, and on the other hand also to reflect how the Web was used in former times. Events such as iPres and IWAW, and consortia such as Netpreserve, as well as various local Web archiving initiatives, tackle this task. There are numerous challenges \cite{masanes2006web} due to the facts that Web pages are ephemeral and due to unpredictable additions, deletions and modifications. Moreover, the hidden Web poses a further challenge. Approaches to Web Archiving are manifold, and range from Web Crawling/Harvesting, server-side archiving, transaction-based archiving, archiving the content of Web databases to library-like approaches advocating persistent identifiers (e.g., Digital Object Identifier - DOI). Especially in the use case about Web database content archiving, Web data extraction techniques are exploited.

Another possibility of archiving the Web is to archive how Web pages are consumed. We can consider this as a type of event-based archiving, that makes sense especially for rich Web applications. The idea is not to archive everything, but to archive selectively sample pathes through an application. This requires to choose sample sites, get an understanding about common and frequent pathes through an application and store the interaction sequence. In a museum-like approach, such selected sequences are stored and can be restored or replayed to provide the user with the experience how the Web was consumed. Deep Web navigation techniques and form understanding are the key technologies here.
In the Social Web scenario, the Blogforever project\footnote{\url{blogforever.eu}} developed some techniques to harvest, preserve, manage and reuse blog content. The information extraction procedure is fully automated and uses Web feeds (i.e., structured XML documents allowing to access the content of a Web page) plus the hypertext of the blog \cite{gkotsis2013self}. Web feeds are used to train and generate a wrapper which is described by a set of simple extraction rules. Retrieved content is finally parsed in such a way as to render the captured content into structured XML data.
\subsubsection{Summary}

\begin{figure}[!ht] \centering
	\includegraphics[width=\columnwidth, trim=0 30 100 0]{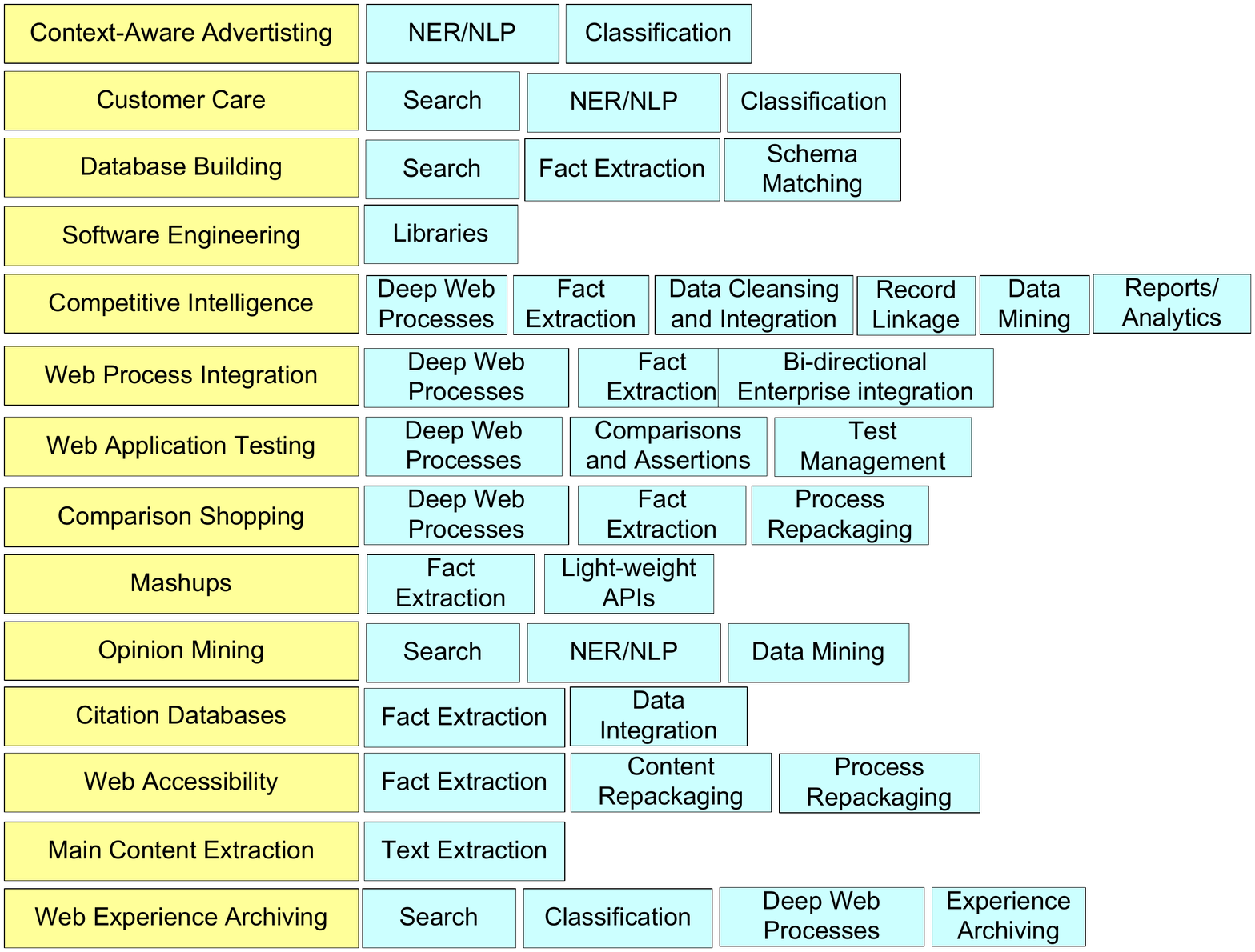}
	\caption{Summary of the application domains in the enterprise context.}
	\label{fig:comparison}
\end{figure}

Figure \ref{fig:comparison} summarizes the 14 discussed enterprise application scenarios. For each scenario we describe the main
ingredients of the value chain, as illustrated in the scenario descriptions.

\subsection{Social Web Applications}
\label{sec:social-applications}

In the latest years, Social Web platforms emerged as one of the most relevant phenomenon on the Web: these platforms are built around users, letting them to create a \textit{web of links between people}, to share thoughts, opinions, photos, travel tips, etc. In such a scenario, often called {\em Web 2.0} users turn from {\em passive consumers} of contents to {\em active producers}.
Social Web platforms provide novel and unprecedented research opportunities. The analysis on a large, often planetary, scale of patterns of users interactions provides the concrete opportunity of answering questions like these: how does human relationships (e.g., friendship relationships) are created and evolve over time \cite{kleinberg2008convergence}? How does novel ideas spread and propagate through the web of human contacts \cite{bettencourt2006power}? How does the human language evolve through social interactions (e.g., how do person expand their lexicon on the basis of their interactions with other persons) \cite{mathes2004folksonomies}?

Besides scientific questions, the analysis of patterns of human interactions in Social Web platforms has also relevant implication at the business level: if we are able to understand the dynamics of interactions among humans, we are also able to identify how groups of users aggregate themselves around shared interests. This is a crucial step for marketing purposes: once users have been grouped, we can, for instance, selectively disseminate commercial advertisements only to those groups formed by users who are actually interested in receiving those advertisements. In an analogous fashion, the fabric of social interactions can be used to identify {\em influential users}, i.e., those users whose commercial behaviors are able to stimulate the adoption/rejection of a given product by large masses of users.

Finally, Social Web users often create accounts and/or profiles in {\em multiple platforms} \cite{Szomszor08,de09finding}. Correlating these accounts and profiles is a key step to understand how the design features and the architecture of a Social Web platform impact on the behavior of a user: so, for instance, one may ask whether some functionalities provided by a given platform augment the aptitude of users to socialize or they impact on the volume of contents produced by a user. Once the relationship between the features of a given platform and the behavior of a user has been elucidated, the designers/managers of that platform can provide novel services to raise the level of engagement of a user in the platform or to raise their degree of loyalty (e.g., to avoid users become inactive in the platform and migrate to other ones).

In the context above, Web Data Extraction techniques play a key role because the capability of timely gathering large amounts of data from one or more Social Web platforms is a an indefeasible tool to analyze human activities. Traditional Web Data Extraction techniques are challenged by new and hard problems both at the technical and scientific level. First of all, Social Web platforms feature a high level of dynamism and variability because the web of contacts of a user or a group of users may significantly vary in small time slots; therefore, we need to design Web Data extraction algorithms/procedures capable of gathering large amount of data in a quick fashion, so that the fragments of collected data keep pace with changes occurring in the structure of the user social network. If such a requirement is not satisfied, the picture emerging from the analysis we can carry out on the data at our disposal could be wrong and it would fail to capture the structure and evolution of human interactions in the platform. A second challenge depends on the fact that Web Data Extraction algorithms are able to capture only a portion of the data generated within one or more platforms. Therefore, we are required to check that the features of a data sample generated as the output of a Web Data Extraction algorithm replicate fairly well the structure of the original data within the platform(s). Finally, since data to gather are associated with humans or reflect human activities, Web Data Extraction techniques are challenged to provide solid guarantees that user privacy is not violated.

In the remaining of this section we will provide an answer to these questions and illustrate how Web Data Extraction techniques have been applied to collect data from Social Web platforms. First of all, we describe how to gather data about user relationships and activities within a single social platform (see Section \ref{sub:socialnetwork}). Secondly, we consider users who created and maintain an account in multiple Social Web platforms and  discuss issues and challenges related to data collection in this scenario (see Section \ref{sub:multiple}).

\subsubsection{Extracting data from a single Online Social Web platform}
\label{sub:socialnetwork}

In this section we first discuss the technological challenges arising if we aim at collecting data about {\em social relationships}, {\em user activities} and the {\em resources} produced and shared by users. Subsequently, we discuss privacy risks associated with the consumption/usage of human related data.

\paragraph*{Technical challenges for collecting data from a single Social Web platform}
\label{subsub:challenges}

We can classify techniques to collect data from a Social Web platform into two main categories: the former category relies on the usage of ad-hoc APIs, usually provided by the Social Web platform itself; the latter relies on HTML scraping.

As for the first category of approaches, we point out that, today, Social Web platforms provide powerful APIs (often available in multiple programming languages) allowing to retrieve in an easy and quick fashion a wide range of information from the platform itself. This information, in particular, regards not only social connections involving members of the platforms but also the content the users posted and, for instance, the tags they applied to label available content.

We can cite the approach of \cite{Kwak10www} as a relevant example of how to collect data from a Social Web platform by means of an API. In that paper, the authors present the results of the crawling of the whole Twitter platform. The dataset described in \cite{Kwak10www} built consisted of 41.7 million user profiles, 1.47 billion social relations; in addition to collecting information about user relationships, the authors gathered also information on tweets and, by performing a semantic analysis, also on the main topics discussed in these tweets. The final dataset contained 4,262 trending topics and 106 million tweets.

From a technical standpoint we want to observe that: {\em (i)} The Twitter API allows to access the whole {\em social graph}, i.e., the graph representing users and their connections in Twitter, without authentication. Other Social Web platforms and the APIs they offer, however, do not generally allow to access the whole social graph. A meaningful example is given by the Facebook API. {\em (ii)} The Twitter API, by default, allows a human user or a software agent to send only 150 requests per hour: this could be an inadmissible limitation because the amount of information generated within Twitter in a relatively small time slot can be very large, and, then, changes in the Twitter network topology could not be properly sensed. To overcome this problem, Twitter offers {\em white lists}: users registered to white lists can send up to 20 000 requests per IP per hour. In \cite{Kwak10www} the authors used a group of 20 computers, each of them belonging to the Twitter white lists, to perform a real-time monitoring of Twitter.

Approaches based on the scraping of HTML pages are able to overcome the limitations above even if they are more complicated to design and implement. To the best of our knowledge, one of the first attempt to crawl large Online Social Networks was performed by Mislove et al. \cite{Mislove2007}. In that paper, the authors focused on platforms like Orkut, Flickr and LiveJournal. To perform crawling, the approach of \cite{Mislove2007} suggests to iteratively retrieve the list of friends of a user which have not yet been visited and to add these contact to the list of users to visit.
According to the language of the graph theory, this corresponds to perform a Breadth-First-Search (BFS) visit of the Social Network graph. The user account from which the BFS starts is often called {\em seed node}; the BFS ends when the whole graph is visited or, alternatively, a stop criterium is met. The BFS is easy to implement and efficient; it produces accurate results if applied on social graphs which can be modeled as unweighted graphs. Due to these reasons, it has been applied in a large number of studies about the topology and structure of Online Social Networks (see, for instance, \cite{Chau2007,Wilson2009,Gjoka2010,Ye2010,Catanese2011}).

As observed by \cite{Mislove2007}, BFS may incur in heavy limitations. First of all, a crawler can get trapped into a strong connected component of the social graph. In addition, if we would use the BFS sample to estimate some structural properties of the social network graph, some properties could be overestimated while others could be underestimated \cite{kurant2010bias}.

To alleviate these problems, several authors suggested more refined sampling techniques. The implementation of these techniques is equivalent to define new Web Data Extraction procedures. Most of them have been discussed and exploited in the context of Facebook \cite{Gjoka2010,Catanese2011} but, unfortunately, some of them can not be extended to other platforms.

In particular, Gjoka et al. \cite{Gjoka2010} considered different visiting algorithms, like BFS, ``Random Walks'' and ``Metropolis-Hastings Random Walks''. A particular mention goes to a visiting method called {\em rejection sampling}. This technique relies on the fact that a truly uniform sample, of Facebook users can be obtained by generating uniformly at random a 32-bit user ID and, subsequently, by polling Facebook about the existence of that ID. The correctness of this procedure derives from the fact that each Facebook user was uniquely identified by a numerical ID ranging from 0 and $2^{32}-1$. Of course, such a solution works well for Facebook but it could not work for other platforms.

In \cite{Catanese2011}, the authors designed a Web Data Extraction architecture based on Intelligent Agents 
(see Figure 1 of \cite{Catanese2011}).
Such an architecture consists of three main components: {\em (i)} a server running the mining agent(s); {\em (ii)} a cross-platform Java application, which implements the logic of the agent; {\em (iii)} an Apache interface, which manages the information transfer through the Web. The proposed architecture is able to implement several crawling strategies like BFS or rejection sampling.

The sampling procedure in \cite{Catanese2011} works as follows: an agent is activated and it queries the Facebook server(s) to obtain the list of Web pages representing the list of friends of a Facebook user. Of course, the Facebook account to visit depends on the basis of the crawling algorithm we want to implement. After parsing this list of pages, it is possible to reconstruct a portion of the Facebook network. Collected data can be converted in XML format so that they can be exploited by other applications (e.g., network visualization tools).




In addition to gathering data about social relationship, we may collect contents generated by users. These contents may vary from resources posted by users (like photos in Flickr or videos in YouTube) to tags applied for labeling resources with the goal of making the retrieval of these resources easier or to increase the visibility of the contents they generated.

\paragraph*{Privacy pitfalls in collecting user related data}
\label{subsub:privacysingle}

The major concern about the extraction of data from Social Web platforms is about {\em privacy}. Several researchers, in fact, showed that we can disclose private information about users by leveraging on publicly available information available in a Social Web platform.
For instance, \cite{hecht2011tweets} provided an approach to finding user location on the basis of user tweets. The authors used basic machine learning techniques which combined user tweets with geotagged articles in Wikipedia. In an analogous fashion, \cite{kinsella2011m} used geographic coordinates extracted from geotagged Twitter data to model user location. Locations were modeled at different levels of granularity ranging from the zip
code to the country level. Experimental studies in \cite{kinsella2011m} show that the proposed model can predict country, state and city with an accuracy comparable with that achieved by industrial tools for geo-localization.

Crandall et al. \cite{crandall2009mapping}  investigated how to organize a large collection of geotagged photos extracted from Flickr.
The proposed approach combined content analysis (based on the textual tags typed by users to describe photos), image analysis and structural analysis based on geotagged data. The most relevant result is that it is possible to locate Flickr photos with a high precision by identifying landmarks via visual, temporal and textual features.

In \cite{chaabane12y} the authors considers a range of user signals expressing user interests (e.g., ``I like'' declaration in Facebook) which are often disclosed by users. By means of a semantic-driven inference technique based on an ontologized version of Wikipedia, the authors show how to discover hidden information about users like gender, relationship status and age.

From the discussion above it emerges that collecting seemingly innocuous data from a Social Web platform hides high privacy risks. Therefore, collected data should be manipulated so that reducing these risks.

Finally, due to privacy settings, the so-called {\em black hole problem} can arise \cite{Ye2010}. In detail, some OSN users may decide to hide their personal information (like contact lists and posts) to strangers and, in this way, they obstacle the crawler in collecting OSN data. \cite{Ye2010} experimentally studies the impact of black holes by considering both different crawling strategies as well as different OSNs (like YouTube and LiveJournal). Black holes have a small {\em but non-negligible} impact on the number of vertices and edges visited by the crawler: on average, vertex/edge coverage decreases in fact between 7\% and 9\% w.r.t. a configuration in which black holes are not present.

\subsubsection{Extracting data from multiple Online Social Web platforms}
\label{sub:multiple}

Social Web users often create and maintain different profiles in different platforms with different goals (e.g., to post and share their photos in Flickr, to share bookmarks on Delicious, to be aware on job proposals on LinkedIn and so on).
We discuss technical approaches to collecting data from multiple Social Web platforms as well as the opportunities coming from the availability of these data. Finally, we describe potential privacy risks related to the management of gathered data.

\paragraph*{Collecting data from multiple Social Web platforms}
\label{subsub:collectingmultiple}

The main technical challenges encountered by Web Data Extraction techniques to collect data from multiple Social Web platforms consists of linking information referring to the same user or the same object. Early approaches were based on ``ad-hoc'' techniques; subsequent approaches featured a higher level of automation and were based on ad-hoc APIs like the Google Social Graph API.

To the best of our knowledge, one of the first approaches facing the problem of correlating multiple user accounts was presented in \cite{Szomszor08}. In that paper, the authors started with a list of 667, 141 user accounts on Delicious such that each account was uniquely associated with a Delicious profile. The same procedure was repeated on a data sample extracted from Flickr. The first stage in the correlation process consisted in comparing usernames in Delicious and Flickr: if these strings exactly match, the two accounts were considered as referring to the same person. In this way, it was possible to build a {\em candidate list} consisting of 232, 391 usernames such that each user name referred to a Flickr and Delicious profile. Of course, such a list must be refined because, for instance, different users may choose the same username. Since both in Delicious and Flickr the users had the chance of filling a form by specifying their real names, the authors of \cite{Szomszor08} suggested to refine the candidate list by keeping only those user accounts whose real names matched exactly. Such a procedure significantly lowered the risk of incurring in false associations but at the same time, dramatically reduced the size of the candidate list to only 502 elements. Some tricks aiming at producing a larger and accurate dataset were also proposed in \cite{Szomszor08}: in particular, the authors observed that, in real scenarios, if a user creates accounts in several Web sites, she frequently adds a link to her accounts: so by using traditional search engines, we can find all pages linking to the homepage of a user and, by filtering these hits we can find the exact URL of the profile of a user in different platforms.

The process described above can be automatized by exploiting {\em ad hoc tool}. Among these tools, the most popular was perhaps the {\tt Google Social Graph API}, even if such an API is no longer available. Such an API is able to find connections among persons on the Web. It can be queried through an HTTP request having a URL called {\em node} as its parameter. The node specifies the URL of a Web page of a user $u$.
The Google Social Graph API is able to return two kinds of results:

\begin{itemize}

\item {\em A list of public URLs that are associated with $u$}; for instance, it reveals the
    URLs of the blog of $u$ and of her Twitter page.

\item {\em A list of publicly declared connections among users}. For instance, it returns the
    list of persons who, in at least one social network, have a link to a page which can be
    associated with $u$.
\end{itemize}

The Google Social Graph API has been used in \cite{Abel*12}.

The identification of the connections between persons or Web objects (like photos or videos) is the key step to design advance services often capable of offering a high level of personalization.
For instance, in \cite{ShLaHa11}, the authors suggest to use tags present in different Social Web systems to establish links between items located in each system. In \cite{Abel*12} the system Mypes is presented. Mypes supports the linkage, aggregation, alignment and semantic enrichment of user profiles available in various Social Web systems, such as Flickr, Delicious and Facebook. In the field of Recommender Systems, the approach of \cite{de09finding} show how to merge ratings provided by users in different Social Web platforms with the goal of computing reputation values which are subsequently used to generate recommendations.


Some authors have also proposed to combine data coming from different platforms but referring to the same object. A nice example has been provided in \cite{Stewart*09}; in that paper the authors consider users of blogs concerning music and users of {\tt Last.fm}, a popular folksonomy whose resources are musical tracks. The ultimate goal of \cite{Stewart*09} is to enrich each Social Web system by re-using tags already exploited in other environments. This activity has a twofold effect: it first allows the automatic annotation of resources which were not originally labeled and, then, enriches user profiles in such a way that user similarities can be computed in a more precise way.

\paragraph*{Privacy risks related to the management of user data spread in multiple platforms}
\label{subsub:privacymultiple}

The discussion above show that the combination and linkage of data spread in independent Social Web platforms provide clear advantages in term of item description as well as on the level of personalization that a system is able to provide to its subscribers. However, the other side of the coin is given by the {\em privacy problems} we may incur when we try to glue together data residing on different platforms.

Some authors recently introduced the so-called {\em user identification} problem, i.e., they studied what information is useful to disclose links between the multiple accounts of a user in independent Social Web platforms.

One of the first approaches to dealing with the user identification problem was proposed in \cite{VoHoSh09}. In that paper the  authors focused on Facebook and StudiVZ ({\tt studivz.net}) and investigated which profile attributes can be used to identify users.
In \cite{zafarani2009connecting}, the authors studied 12~different Social Web systems (like Delicious, Flickr and YouTube) with the goal of finding a mapping involving the different user accounts. This mapping can be found by applying a traditional search engine.
In \cite{Iofciu*11}, the authors suggest to combine profile attributes (like usernames) with an analysis of the user contributed tags to identify users. They suggest various strategies to compare the tag-based profiles of two users and some of these strategies were able to achieve an accuracy of almost 80\% in user identification.
\cite{perito2011unique} explored the possibility of linking users profiles only by looking at their usernames. Such an approach is based on the idea that the probability that two usernames are associated with the same person depends on the entropies of the strings representing usernames.

\cite{Balduzzi*10} described a simple but effective attack to discover the multiple digital identities of a user. Such an attack depends on the fact that a user often subscribes to multiple Social Web platforms by means of a {\em single e-mail address}. An attacker can query Social Web platforms by issuing a list of e-mail addresses to each Social Web platform; once the profiles of a user have been identified in each platform the data contained in these profiles can be merged to obtain private personal information. The authors considered a list of about 10.4 millions of e-mail addresses and they were able to automatically identify more than 1.2 millions of user profiles registered in different platforms like Facebook and XING. The most popular providers acknowledged the privacy problem raised in \cite{Balduzzi*10} and implemented ad hoc countermeasures.

An interesting study is finally presented in \cite{goga2012exploiting}. The authors showed that correlating seemingly innocuous attributes describing user activities in multiple social networks allow attackers to track user behaviors and infer their features. In particular, the analysis of \cite{goga2012exploiting} focused on three features of online activity like the geo-location of users posts, the timestamp of posts, and the user's writing style (captured by proper language models).

\subsection{Opportunities for cross-fertilization}
\label{sec:cross}

In this section we discuss on the possibility of re-using Web Data Extraction techniques originally developed in a given application domain to another domain. This discussion is instrumental in highlighting techniques which can be applied across different application domains and techniques which require some additional information (which can be present in some application domains and missing in others).

In the most general case, no assumption is made on the structure and content of a collection of Web pages which constitute the input of a Web Data Extraction tool. Each Web page can be regarded as a text document and, in such a configuration, approaches relying on Regular Expressions can be applied independently of the application domain we are considering. As mentioned before, regular expressions can be regarded as a formal language allowing to find strings or patterns from text according to some matching criteria. Approaches based on regular expressions therefore may be suitable when we do not have, at our disposal, any information about the structure of the Web pages. Of course, these approaches can be extended so that to take some structural elements into account (like HTML tags). The usage of regular expression may be disadvantageous if need to collect large collections of documents and assume that those documents deal with different topics. In such a case, we need complex expressions for extracting data from documents and this would require a high level of expertise.

The next step consists of assuming that some information on the structure of the document is available. In such a case, wrappers are an effective solution and they are able to work fairly well in different application domains. In fact, the hierarchical structure induced by HTML tags associated with a page often provides useful information to the Web Data Extraction task. A powerful solution taking advantage of the structure of HTML pages derives from the the usage of XPath to quickly locate an element (or multiple instances of the same element in the document tree). Approaches based on XPath have been first exploited in the context of Enterprise applications and, later, they have been successfully re-used in the context of Social Web applications. There are, however, a number of requirements an application domain should satisfy in order to enable the usage of XPath. First of all, the structure of Web pages should be perfectly defined to the wrapper designer (or to the procedure inducing the wrapper); such an assumption, of course, can not be true in some domains because data are regarded as {\em proprietary} and technical details about the structure of the Web wrappers in that domain are not available. In addition, we should assume a certain {\em level of structural coherence} among all the Web pages belonging to a Web source. Such an assumption is often true both in enterprise domain and in many Social Web systems: in fact, Web pages within an organization (e.g., a company) or a Social Web platform often derive from the same template and, therefore, a certain form of structural regularity across all the pages emerge. By contrast, if we plan to manage pages with different structures (even if referring to the same domain), we observe that even small changes in the structure of two Web pages may have a devastating impact and thus would require to entirely rewrite the wrapper.

A nice example of cross-fertilization is in the context of the crawling of Online Social Networks \cite{Catanese2011}. In that paper the authors implemented a Web wrapper to crawl Facebook largely exploiting techniques and algorithms which were part of the Lixto suite and that were originally designed to work for Business and Competitive Intelligence applications. The proposed wrapper is based on the execution of XPath queries; human experts, in the configuration phase, are allowed to specify what elements have to be selected.


The crawler provides two running modes: {\em (i)} visual extraction and, {\em (ii)} HTTP request-based extraction. In the visual extraction mode,
the crawler embeds a Firefox browser interfaced through XULRunner ({\tt developer.mozilla.org/en/XULRunner}) via XPCOM. The visual approach requires the rendering of the Web pages which is a time-consuming activity. Therefore, to extract large amounts of data, \cite{Catanese2011} suggested to send HTTP requests to fetch Web pages.

In some application scenarios, in addition to assuming that the syntactic structure of a Web page is known, it is possible to assume that a rich semantic structure emerges from the Web pages. If such an hypothesis holds true, techniques from Information Extraction (IE) and Natural Language Processing (NLP) can be conveniently used \cite{54,53,50}. The range of applications benefiting from NLP techniques comprises relevant examples both in the enterprise and Social Web scenarios: for instance, relevant applications of NLP/IE techniques are the extraction of facts from speech transcriptions in forums, email messages, newspaper articles, resumes etc.

\section{Conclusions}
\label{sec:conclusions}

The World Wide Web contains a large amount of unstructured data. The need for structured information urged researchers to develop and implement various strategies to accomplish the task of automatically extracting data from Web sources. Such a process is known with the name of Web Data Extraction and it has had (and continues to have) a wide range of applications in several fields, ranging from commercial to Social Web applications.

The central thread of this survey is to classify existing approaches to Web Data Extraction in terms of the applications for which they have been employed.

In the first part of this paper, we provided a classification of algorithmic techniques exploited to extract data from Web pages. We organized the material by presenting first basic techniques and, subsequently, the main variants to these techniques. Finally, we focus on how Web Data Extraction systems work in practices. We provide different perspectives to classify Web Data Extraction systems (like the ease of use, the possibility of extracting data from the Deep Web and so on).

The second part of the survey is about the applications of Web Data Extraction systems to real-world scenarios. We provided a simple classification framework in which existing applications have been grouped into two main classes (Enterprise and Social Web applications). The material has been organized around the application domains of Web Data Extraction systems: we identify, for each class, some sub-domains and described how Web Data Extraction techniques work in each domain. This part ends with a discussion about the opportunities of cross-fertilization.



We discuss now some possible future (and, in our opinion, promising) applications of Web Data Extraction techniques.

\paragraph*{Bio-informatics and Scientific Computing}
\label{sub:bioinformatics}

A growing field of application of Web Data Extraction is \textit{bio-informatics}: on the World Wide Web it is very common to find medical sources, in particular regarding \textit{bio-chemistry} and \textit{genetics}.
Bio-Informatics is an excellent example of the application of scientific computing -- refer e.g.\ to \cite{68} for a selected scientific computing project.

Plake et al. \cite{14} worked on PubMed ({\tt www.pubmed.com}) -- the biggest repository of medical-scientific works that covers a broad range of topics -- extracting information and relationships to create a graph; this structure could be a good starting point to proceed in extracting data about \textit{proteins} and \textit{genes}, for example connections and interactions among them: this information can be usually found, not in Web pages, rather they are available in PDF or Postscript format.
In the future, Web Data Extraction should be extensively used also to these documents: approaches to solving this problem are going to be developed, inherited, both from Information Extraction and Web Data Extraction systems, because of the semi-structured format of PostScript-based files.
On the other hand, web services play a dominant role in this area as well, and another important challenge is the intelligent and efficient querying of Web services as investigated by the ongoing SeCo project  ({\tt www.search-computing.it}).

\paragraph*{Web harvesting}
\label{sub:webharvesting}

One of the most attractive future applications of the Web Data Extraction is \textit{Web Harvesting} \cite{29}: Gatterbauer \cite{13} defines it as ``the process of gathering and integrating data from various heterogeneous Web sources''.
The most important aspect (although partially different from specific Web Data Extraction) is that, during the last phase of data transformation, the amount of gathered data is \textit{many times greater} than the extracted data.
The work of filtering and refining information from Web sources ensures that extracted data lie in the domain of interest and are relevant to users: this step is called \textit{integration}.
The Web harvesting remains an open problem with large margin of improvement: because of the billions of Web pages, it is a computational problem, also for restricted domains, to crawl enough sources from the Web to build a \textit{solid ontological base}.
There is also a human engagement problem, correlated to the degree of automation of the process: when and where humans have to interact with the system of Web harvesting?
Should be a fully automatic process? What degree of precision can we accept for the harvesting? All these questions are still open for future works.
Projects such as the DIADEM ({\tt web.comlab.ox.ac.uk/projects/DIADEM}) at Oxford University tackle the challenge for fully automatic generation of wrappers for restricted domains such as real estate.

\paragraph*{Linking Data from several Web sources}
\label{sub:linking}

A growing number of authors suggest to integrate data coming from as many sources as possible to obtain a detailed description of an object. Such a description would be hard to obtain if we would focus only on a single system/service.

The material discussed in this section partially overlaps with some ideas/techniques presented in Section \ref{sub:multiple}. However, in Section \ref{sub:multiple} we focused on Social Web platforms and showed that linking information stored in the profiles of a user spread across multiple platforms leads to a better identification of user needs and, ultimately, raises the level of personalization a system can offer to her. In this section, by contrast, we focus on systems having two main features. {\em (i)} they often publicly expose their data on the Web and {\em (ii)} these systems are {\em lacking} of a social connotation, i.e., they do not target at building a community of interacting members.

A first research work showing the benefits of linking data provided by independent Web systems is provided in \cite{szomszor2007folksonomies}. In that paper, the authors combined information from the Internet Movie Database ({\tt www.imdb.com}) and Netflix ({\tt www.netflix.com}). The IMDB is an online database containing extensive information on movies, actors and television shows. IMDB users are allowed to add
tags to describe the main features of a movie (e.g., the most important scenes, location, genres and so on). Netflix is an US-based company offering an online DVD rental service. Netflix users are allowed to rate a movie by providing a score. In \cite{szomszor2007folksonomies}, data from Netflix and IMDB were imported in a relational DBMS; movie titles in IMDB were correlated with movie titles in Netflix by applying string matching techniques. In this way, for each movie, the Netflix ratings and the IMDB description were available. The authors studied three recommendation strategies based on ratings alone, on tags alone and, finally, a combination of ratings and tags. Experimental trials provided evidence that the combination of data located in different systems was able to improve the level of accuracy in the provided recommendations. Spiegel et al. presented an analogous study in \cite{SpKuLi09}; in that paper the authors combine user ratings (coming from MovieLens database) with movie information provided by IMDB.

The linkage of datasets coming from independent Web platforms fuels novel scientific applications. One of these application is given from {\em cross-domain recommender systems}, i.e., on recommender systems running on multiple domains. For example, one can use user ratings about movies to predict user preferences in the music domain. In a given domain (e.g., if our main aim is to recommend movies), information about user preferences may be insufficient and, in many case, most of this information is missing for a non-negligible part of user population. By contrast, a relevant amount of information describing user preferences could be available (at a limited cost) in other domains. We could {\em transfer} this information from a domain to another one with the goal of dealing with data sparsity and improving recommendation accuracy.

Various strategies have been proposed to perform such a transfer of information and some of these strategies take advantages from models and algorithms developed in the field of {\em transfer of learning} \cite{pan2010survey}. For instance, some approaches use co-clustering \cite{li2009can}, other clustering techniques in conjunction with probabilistic models \cite{li2009transfer} and, finally, other approaches project the user and item space in the various domains in a shared latent space by means of regularization techniques \cite{zhang2010multi}.

\paragraph*{Scalability Issues and the usage of Cloud Computing infrastructures}. Because of the increasing complexity of Web Data Extraction procedures, some authors suggested to use Cloud computing services like Amazon EC2 to provide a high level of reliability and scalability.
In cloud architectures, computing resources can be accessed as services which are paid by end users depending on the usage time or on the amount of data to process. One of the first Web Data Extraction platforms relying on a cloud architecture is Lixto \cite{62}: in it cloud clients receive a wrapper as well as extraction parameters and they pass back retrieved data; these data are not stored on cloud instances.
Cloud architectures provide intelligent procedures to fairly (and intelligently) distribute the computational load amongst the available processing units and this yields a more efficient resource management. Due to this reasons we believe that in the next years a growing number of Web Data Extraction platforms will rely on cloud services.

\paragraph*{Extracting Web Content into Semantic Web format}. The popularity of the open linked data initiative prompted some authors to develop systems which support the extraction of Web content and their storage in some Semantic Web format. Among these systems we cite the above mentioned DEiXTo system, which DEiXTo which to build extraction rules capable of converting extracted content in any structured format (and, among them, RDF). Other related projects are Virtuoso Sponger\footnote{\url{http://virtuoso.openlinksw.com/dataspace/doc/dav/wiki/Main/VirtSponger}}, which
generates Linked Data starting from different data sources and supports a wide range of data representation formats, Semantic Fire\footnote{\url{https://code.google.com/p/semantic-fire/}}, which allows to extract and store the data from a Web site as RDF.

\bibliography{web-biblio}
\bibliographystyle{abbrv}

\end{document}